\documentclass[preprint,12pt]{elsarticle}

\usepackage{graphicx}
\usepackage{subcaption}
\usepackage{multirow}
\usepackage{array}
\usepackage{amsmath,amssymb,amsfonts}
\usepackage{amsthm}
\usepackage{mathrsfs}
\usepackage[title]{appendix}
\usepackage{textcomp}
\usepackage{manyfoot}
\usepackage{siunitx}
\usepackage{booktabs}
\usepackage{algorithm}
\usepackage{algorithmicx}
\usepackage{algpseudocode}
\usepackage{listings}
\usepackage{enumitem}
\usepackage{hyperref}
\usepackage[version=4]{mhchem}
\usepackage{float}
\usepackage{orcidlink}
\usepackage{makecell}

\begin{document}

\begin{frontmatter}
\title{Machine Learning for Cloud Detection in IASI Measurements: A Data-Driven SVM Approach with Physical Constraints}

\author[3,1]{Chiara Zugarini \orcidlink{0009-0005-8053-0646} \corref{cor1}}
\ead{chiara.zugarini@unifi.it}

\author[2]{Cristina Sgattoni \orcidlink{0000-0001-5734-0856}}
\ead{cristina.sgattoni@cnr.it}

\author[1]{Luca Sgheri \orcidlink{0000-0002-6014-9363}}
\ead{luca.sgheri@cnr.it}

\cortext[cor1]{Corresponding author. Email: chiara.zugarini@unifi.it}

\affiliation[1] {organization={CNR-IAC},
                 addressline={Via Madonna del Piano, 10},
                 city={Sesto Fiorentino},
                 postcode={I-50019},
                 state={FI},
                 country={Italy}
}
\affiliation[2] {organization={CNR-IBE},
                 addressline={Via Madonna del Piano, 10},
                 city={Sesto Fiorentino},
                 postcode={I-50019},
                 state={FI},
                 country={Italy}
}
\affiliation[3] {organization={Università di Firenze, Dipartimento di Ingegneria Informatica},
                 addressline={Via di S. Marta, 3},
                 city={Firenze},
                 postcode={I-50139},
                 state={FI},
                 country={Italy}
}

\begin{abstract}
Cloud detection is a fundamental step for the physical interpretation and 
operational exploitation of hyperspectral infrared sounders, yet the 
extent to which cloud information can be retrieved from infrared radiances 
alone has not been fully assessed. Here we introduce the Cloud 
Identification Support Vector Machine (CISVM), a supervised framework for 
global clear and cloudy classification from Infrared Atmospheric Sounding 
Interferometer (IASI) Level 1C observations.

The analysis spans four seasons and compares different input representations (radiances and brightness temperatures), spectral reductions (cloud-sensitive channel subsets and PCA), and stratifications by surface type and climate zone. The AVHRR-derived IASI cloud cover product provides the supervisory labels, while the trained classifier operates exclusively on hyperspectral infrared radiances. Its performance is evaluated on an independent IASI test set and compared with collocated MODIS cloud products, providing an agreement assessment with operational references rather than an absolute validation of cloud detection accuracy.

The best-performing configuration combines radiances with PCA and reaches 88.52\% agreement with the operational IASI cloud reference. Although the classifier is data-driven, its behaviour remains physically interpretable because the learned statistical relationships consistently reflect the radiative properties of the observed scenes. Performance degradation is concentrated over land and polar environments, where reduced cloud surface radiative contrast limits spectral separability. The soil type and seasonal analyses further indicate that classification uncertainty is structured rather than random, and is strongly governed by surface radiative properties and their seasonal modulation. Overall, these results show that infrared cloud screening can be both operationally promising and scientifically informative, with direct relevance for future hyperspectral sounders equipped with only limited imaging capability, including Far-infrared Outgoing Radiation Understanding and Monitoring (FORUM), the ninth European Space Agency (ESA) Earth Explorer Mission.
\end{abstract}


\end{frontmatter}

\section{Introduction\label{sec:introduction}}
Cloud cover is one of the most critical factors in atmospheric retrieval operations, as it significantly affects the quality and accuracy of atmospheric reconstruction models. Accurate cloud detection is essential not only for understanding atmospheric processes but also for improving climate and weather forecasting. Clouds play a fundamental role in Earth's energy budget by both reflecting incoming solar radiation (producing a cooling effect) and trapping outgoing longwave radiation (producing a warming effect).

Cloud detection can be framed as a classification problem, making it well suited for Machine Learning (ML) approaches. In this study, we employ the Support Vector Machine (SVM)~\cite{vapnik64, cortes1995, vapnik98, burges1998}, a well-established ML technique for classification, which we combine with kernel methods to effectively handle non-linearly separable data.

Our experiments are based on radiances acquired by the Infrared Atmospheric Sounding Interferometer (IASI) onboard the MetOp-C satellite~\cite{blumstein2004, eumetsat-iasi}. Scenes are classified according to the operational IASI Level 1C Cloud Cover (CC) product~\cite{eumetsat-iasi, eumetsat2011iasi}, which provides the percentage of cloud coverage within the instrument's Field of View (FoV). Because this operational reference is derived from collocated Advanced Very High Resolution Radiometer (AVHRR) visible and near infrared measurements, our study specifically investigates the capability to recover cloud information consistent with this multispectral-derived mask using strictly hyperspectral infrared information.

The classifier operates exclusively on IASI infrared radiances. AVHRR measurements are therefore required only during training and are never used during inference. This decoupling between the training reference and the operational input is particularly valuable in observational settings where visible and near-infrared information may be unavailable or less reliable, such as nighttime acquisitions or persistent polar darkness, and more generally for future hyperspectral infrared missions lacking a co-located multispectral imager. Although classification performance in these regimes is somewhat reduced relative to temperate and tropical latitudes, as shown later in this study, the proposed framework remains a promising approach for infrared cloud screening.

To further investigate how surface conditions affect the separability between clear and cloudy scenes, we also incorporate soil-type information from the ERA5 reanalysis dataset~\cite{era5}, produced by the European Centre for Medium-Range Weather Forecasts (ECMWF), enabling data partitioning by soil type and climate zone based on the observation geolocation. This subdivision is therefore intended as a practical stratification of surface radiative regimes rather than a physical description of land cover.

We evaluated several preprocessing strategies, including alternative radiometric representations (radiances and brightness temperatures, BT) and spectral dimensionality reduction via either Principal Component Analysis (PCA) or selection of cloud-sensitive channels.

To comprehensively assess the impact of seasonality and varying atmospheric conditions on classification performance, all experiments are performed across four seasonal periods (Winter, Spring, Summer and Autumn). 

Because the model is trained against the operational AVHRR derived mask, this primary evaluation represents a consistency assessment with the operational reference rather than an absolute ground truth validation. To provide an external consistency check, we also compare our classification results with cloud information derived from the Moderate-Resolution Imaging Spectroradiometer (MODIS) instrument~\cite{justice2002}, with particular attention to high-latitude and polar regions where infrared cloud detection is notoriously challenging.

The contribution of this work is therefore not the introduction of a new machine-learning algorithm, but the demonstration that hyperspectral infrared radiances alone contain sufficient information to reproduce the large-scale behaviour of an operational multispectral cloud product while revealing physically interpretable relationships between classification performance and surface radiative properties. By evaluating classifier complexity, computational costs, and geographical limitations over different surface types, we provide an operational perspective, including an explicit characterization of its regional limitations, that is informative for future satellite missions equipped with highly resolving infrared sounders but lacking advanced multispectral imagers, such as the FORUM mission.

The paper is organized as follows. Section~\ref{sec:background_and_theory} provides general background on cloud detection from satellite data, outlining both traditional and machine learning-based approaches. Section~\ref{sec:training_test_sets} describes the construction of the training and test datasets from IASI L1C measurements, including the adopted thresholds and subdivisions. Section~\ref{sec:svm_to_IASI} presents the classification results obtained using SVM, evaluating the impact of different input representations (radiances and brightness temperatures) and dimensionality reduction strategies, including PCA. Section~\ref{sec:cloud_mask_and_MODIS_data} compares the SVM-based classifications with an independent cloud mask derived from MODIS observations. Finally, Section~\ref{sec:conclusions} summarizes the main outcomes of the study and discusses possible future developments.

Due to the length of the paper, many technical details have been moved to the Supplementary Material. References to its sections use the prefix S (e.g., Section S1).

\section{Background}\label{sec:background_and_theory}

Accurate cloud detection from satellite observations is fundamental not only for preprocessing atmospheric parameters but also for numerical weather prediction and climate monitoring. Since the 1990s, operational algorithms based on multispectral thresholding, adaptive to surface type and illumination conditions, have been developed \cite{Lavanant1999}, achieving accuracy above 92\% in mid-latitude regions.

Wavelet-based methods have been proposed as an auxiliary tool for cloud detection and classification from satellite images~\cite{Amato2001}. By applying 2D discrete wavelet transforms and discriminant analysis, the approach takes advantage of the spatial scale differences between clouds and surface features, enabling effective separation in the wavelet domain. However, no quantitative validation metrics were reported to assess classification performance.

McNally and Watts~\cite{McNally2003} introduced a physical approach based on the analysis of first-guess departures between observations and clear-sky simulations. This method allowed for the identification of clear channels even in partly cloudy scenes, with residual contamination below 0.2~\si{K}.

The Bayesian approach is another method for cloud detection. It was applied to Advanced Very High Resolution Radiometer (AVHRR) data by Heidinger~\cite{Heidinger2012}. The Pathfinder Atmospheres - Extended (PATMOS-x) dataset shows good agreement with MODIS products, despite differences in sensors and processing techniques. This suggests that different methods can still produce consistent cloud fraction estimates.

Stubenrauch~\cite{stubenrauch2013} made an inter-comparison of cloud products obtained from different sensors, highlighting inter-instrument biases.

Supervised classification techniques such as Cumulative Discriminant Analysis (CDA) have been applied to IASI radiances to distinguish between cloudy and clear-sky observations in the thermal infrared \cite{Amato2014}. The method used nine spectral statistics and a Principal Component Analysis (PCA) preprocessing step, achieving more than 80\% agreement with AVHRR and Spinning Enhanced Visible Infra-Red Imager (SEVIRI) cloud masks on global and regional datasets.

The Cloud Identification and Classification (CIC) algorithm~\cite{Maestri2019}, based on PCA, is a machine learning method for cloud detection and scene classification using a univariate distribution and a similarity index threshold. It was tested on a large synthetic dataset covering a wide range of climatic conditions in the framework of FORUM End-to-End simulator project~\cite{sgheri2022}. The method showed high detection rates for clear/cloudy scenes, particularly for thin cirrus clouds, and performance improved significantly when using far-infrared radiances, reaching detection scores up to 90\%. However, performance tends to decrease in polar regions because of the reduced radiative contrast between clouds and the surface.

Supervised neural networks have also been used successfully: Mastro et al. \cite{Mastro2020} achieved 93\% accuracy on IASI L1C spectral radiances, covering the period January 2016 to November 2016. The analysis was limited to specific geographic regions: Eastern Europe and tropical areas, where truth data were taken from a cloud mask product of the AVHRR. The methodology used a multilayer feedforward neural network trained on a subsampled dataset with well-known sky conditions, using PCA and regularization techniques to improve performance.

Recently, Whitburn et al. \cite{Whitburn_cloud_mask} used 45 selected IASI channels and reached 87\% agreement with the operational IASI L1 radiances and L2 cloud product, creating a daily cloud mask using a supervised neural network (NN). The method was designed to ensure temporal consistency across the IASI time series and different Metop platforms, avoiding channels affected by major absorption lines.

In the global context of using SVM for cloud detection, the application by Lee et al. \cite{Lee2004} is particularly relevant. In their work, they applied Multicategory SVM (MSVM) to several simulated MODIS radiance channels, demonstrating the potential of this technique for satellite-based cloud classification, with the best recognition rate corresponding to a misclassification rate of 10.16\%. Although the tests were limited in number and did not cover the entire globe, they clearly highlight the promise of this approach.

In the 2012 study by Addesso et al.~\cite{Addesso2012}, SVM were applied to cloud detection using multispectral images from the SEVIRI sensor. The method employed five spectral channels and introduced a penalty map to model spatial dependencies between pixels, enhancing classification performance. Their improved SVM-A approach achieved an overall accuracy of about 88.3\%, demonstrating significant gains over classical SVM by reducing missed detections.

In the study by Murino et al.~\cite{Murino2014} in 2014, SVM was applied for cloud detection using multispectral MODIS data, including synthetic, expert-annotated, and real datasets. The SVM utilized multispectral radiance features to classify pixels, achieving success rates between 97.1\% and 98.9\% over land, and 98.3\% to 99.2\% over water. The study demonstrated that SVM, alongside k-Nearest Neighbors, provides high performance in cloud mask estimation from MODIS imagery.

SVM has been effectively applied to classify cloud and land categories~\cite{Wohlfarth2018} using multispectral Landsat 8 imagery~\cite{landsat8_nasa}. Training the SVM with both color and thermal channels achieved a high overall accuracy of 95.4\%, highlighting the critical role of thermal radiance information in improving classification performance. These results demonstrate the strength of SVMs in leveraging radiance data for accurate cloud detection.

Maestri et al.~\cite{Maestri2019_refirpad} in 2019 applied SVM to downwelling far-infrared radiance measured by Radiation Explorer in the Far InfraRed Prototype for Applications and Developments (REFIR-PAD) over Antarctica, focusing on thin ice cloud classification. Unlike IASI, which observes upwelling radiance globally in the mid-IR from space, REFIR-PAD provides high-resolution ground-based measurements in the FIR. Their method relied on brightness temperature differences in the [380,575]~\si{cm^{-1}} range, emphasizing local cloud microphysics rather than global-scale retrieval.

To the best of our knowledge, no direct applications of SVM to IASI L1C spectral radiances have been reported in the literature. This study introduces the use of SVM for binary clear/cloudy classification from IASI L1C radiances, using sample subdivisions based on soil type and geographic location of the observed radiances. Future work will focus on extending the method to multiclass cloud classification (ice, water, and mixed-phase clouds), leveraging the generalization capabilities of SVMs, or different techniques, in global-scale applications.

Cloud coverage information is available in the IASI Level 1C products \cite{eumetsat-iasi, eumetsat2011iasi}, derived from the collocated AVHRR 1B radiometer. This parameter, expressed as a percentage, indicates the fraction of the IASI FoV obscured by clouds.
The objective of this study is to develop a machine learning model capable of distinguishing clear and cloudy scenes directly from radiances, without relying on external instruments.

To better understand the classification approach adopted in this study, we briefly review the theoretical foundations of Support Vector Machines.

\section{Creation of Training and Test Sets from IASI Measurements}\label{sec:training_test_sets}

IASI~\cite{blumstein2004} is a Fourier transform spectrometer onboard the MetOp satellites operated by the European Organisation for the Exploitation of Meteorological Satellites (EUMETSAT) and developed in cooperation with the European Space Agency (ESA), designed for atmospheric sounding in the infrared domain. It covers a swath of approximately 2200~\si{km}, corresponding to observation angles up to $\pm 48.3^{\circ}$ from the nadir, enabling near-global coverage with successive orbits.

IASI is currently operated on three successive MetOp satellites (MetOp-A, launched in 2006 and decommissioned in 2021; MetOp-B, launched in 2012; and MetOp-C, launched in 2018), each carrying a nominally identical instrument with consistent spectral characteristics and calibration procedures. All IASI observations used in this study, including both the training dataset and the independent test datasets, were acquired from the same MetOp-C platform, ensuring full consistency of instrument characteristics, spectral response, and calibration between the training and evaluation phases.

IASI provides 8461 spectral channels across its full range, with a nadir spatial resolution (FOV diameter) of approximately 12 km, and measures radiances in the spectral range 645–2760~\si{cm^{-1}} with a spectral resolution of 0.5~\si{cm^{-1}}, and a spectral sampling of 0.25~\si{cm^{-1}}. This high resolution allows for accurate retrieval of vertical profiles of atmospheric temperature, humidity, and trace gases, essential for both weather forecasting and climate monitoring. The instrument also provides data sensitive to cloud cover and surface characteristics, making it particularly valuable for atmospheric composition and pollution studies.

The algorithm is tested using the IASI-L1 estimate. The cloud cover derived from AVHRR is normalized to the $[0,1]$ range, and a threshold of 0.1 is applied to distinguish clear (cloud fraction $< 0.1$) and cloudy (cloud fraction $\geq 0.1$) scenes. We adopt a threshold of 0.1, rather than the stricter zero-cloud-fraction criterion used by Whitburn et al.~\cite{Whitburn_cloud_mask} to define clear-sky reference scenes, in order to retain a sufficiently large and balanced clear-sky sample across all seasons, surface types, and climate zones; a stricter threshold would substantially reduce the size of the clear-sky class, particularly over regions with persistent partial cloudiness. This threshold is consistent with the operational clear and cloudy interpretation adopted throughout the IASI L1C Cloud Cover product, ensuring that the supervised labels used for training and the IASI L1C reference used for comparison share the same clear and cloudy definition; the robustness of this choice is further assessed empirically later in this study. As previously stated, we construct four separate datasets corresponding to the four seasons: March–April for Spring, June–July for Summer, September–October for Autumn, and December–January for Winter.

\subsection{IASI training set}\label{subsec:IASI_dataset}
\begin{figure}[htbp]
\centering
\includegraphics[width=0.95\linewidth]{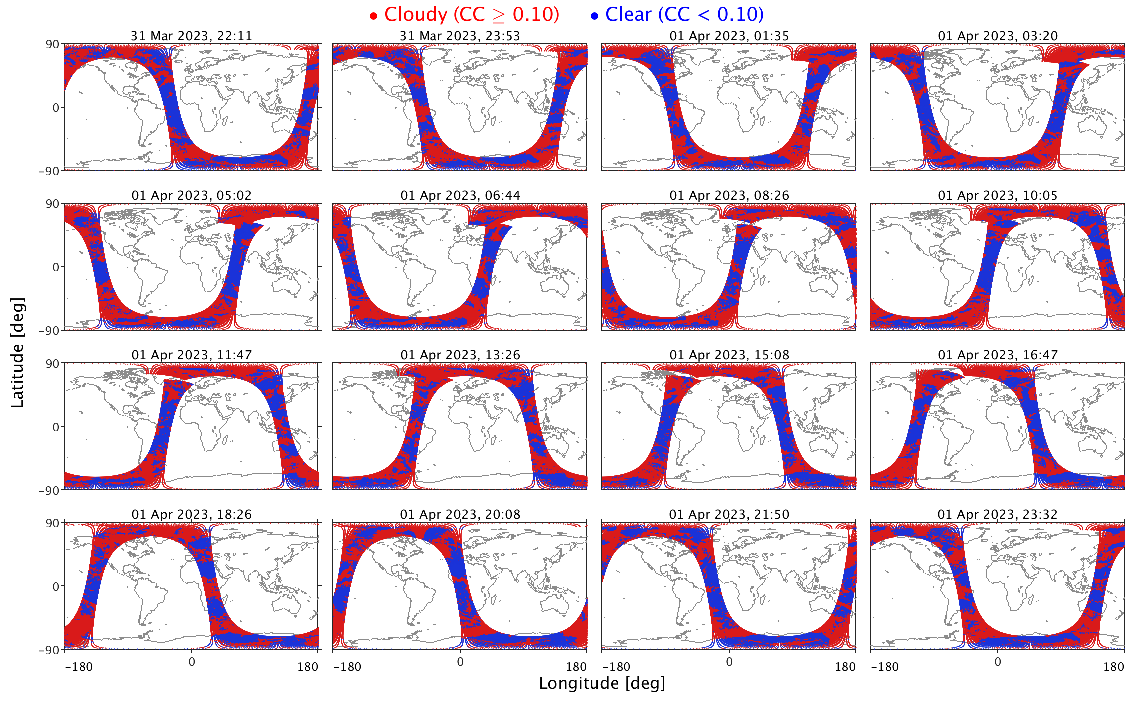}
\caption{Spatial distribution of IASI clear-sky (blue) and cloudy (red) observations for the test set Spring season, aggregated over the 16 corresponding test orbits.}
\label{fig:IASI_cle_clo_trainingset}
\end{figure}

The training dataset is constructed from IASI observations collected in March, April, June, July, September, October, and December 2023, and January 2024. For each season, 16 consecutive daily orbits were selected, covering approximately one day of observations and ensuring near-global coverage; the complete list of the 64 selected orbits, together with the corresponding start/end dates, times, and orbit ID, is reported in Table~S1. The spatial distribution of clear and cloudy sky IASI observations for the Spring season, aggregated over all 16 corresponding orbits, is shown in Figure~\ref{fig:IASI_cle_clo_trainingset}. The corresponding distributions for Summer, Autumn, and Winter are provided in the Figure~S1,~S2,~S3.

\subsection{IASI test set}\label{subsec:IASI_testset}
\begin{figure}[htbp]
\centering
\includegraphics[width=0.95\linewidth]{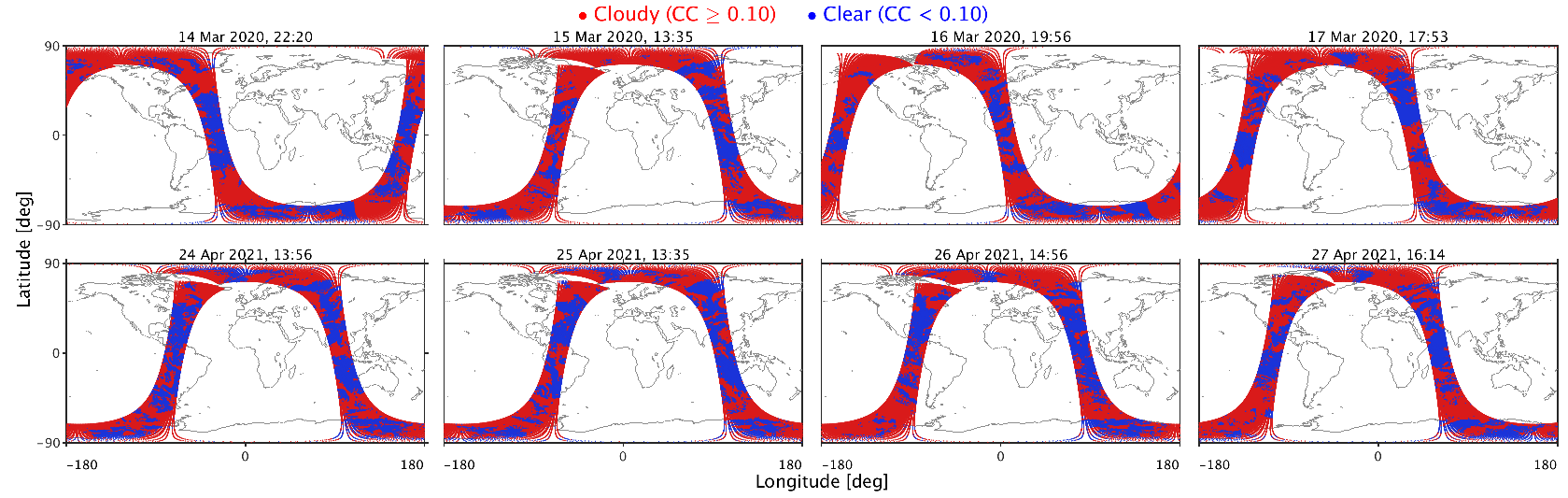}
\caption{Spatial distribution of IASI clear-sky (blue) and cloudy (red) observations for the test set Spring season, aggregated over the 8 corresponding test orbits.}
\label{fig:IASI_cle_clo_testset}
\end{figure}

Test sets are prepared following the same criteria, using 8 orbits per season from the same months as the training dataset, but from previous years (2019-2021), for a total of 32 test orbits. Seasonal subsets (Spring, Summer, Autumn, and Winter) are defined consistently with the training dataset; the complete list of selected orbits, dates, and starting times is reported in Table~S2. Figure~\ref{fig:IASI_cle_clo_testset} shows the spatial distribution of clear and cloudy scenes for the Spring season, aggregated over the 8 corresponding test orbits. The corresponding distributions for Summer, Autumn, and Winter are provided in Figure~S4,~S5,~S6.

\subsection{Additional information and sample subdivisions}\label{subsec:IASI_additional}
In addition to the spectra, we add auxiliary surface information. The soil type is provided as a categorical integer variable representing predefined soil texture classes in the ECMWF land surface dataset~\cite{era5}. The different soil types are shown in Table~\ref{tab:SoilType}.

\begin{table}[H]
    \centering
    \begin{tabular}{|c|c|c|c|}
    \hline
    \textbf{N.} & \textbf{Soil Type} & \textbf{N.} & \textbf{Soil Type} \\
    \hline
    \hline
     0 & Non-land pixels & 4 &  Fine \\
    \hline
     1 & Coarse & 5 &  Very Fine \\
    \hline
     2 &  Medium & 6 &  Organic \\
    \hline
     3 &  Medium fine & 7 &  Tropical Organic \\
    \hline
    \end{tabular}
    \caption{
Description of soil type classes used in this work, based on ERA5 data.}
    \label{tab:SoilType}
\end{table}

\noindent Soil type, rather than a land-cover classification, was adopted as the primary surface descriptor: it is directly available within the same ERA5 dataset already used for geolocation and auxiliary variables, and it remains static across the four seasons considered in this study, ensuring full consistency of spatial resolution and surface classification across all training and test periods. While land-cover classes (e.g., forest, grassland, urban areas) may more directly reflect vegetation and land-use patterns, soil type provides a stable, season-independent surface descriptor strongly linked to emissivity, which directly affects the radiance or brightness-temperature signal exploited by the classifier. 

Water is assigned to a dedicated category of Soil Type 0, as "Non-land pixels". This choice allows the classification performance over water to be evaluated consistently with the land-surface subdivisions, and is further exploited in Section~\ref{subsec:surface_classification_performance} where land and non-land surfaces are compared directly in Figure~\ref{fig:land_nonland_analysis}.

\begin{figure}[htbp]
\centering
\includegraphics[width=1\textwidth]{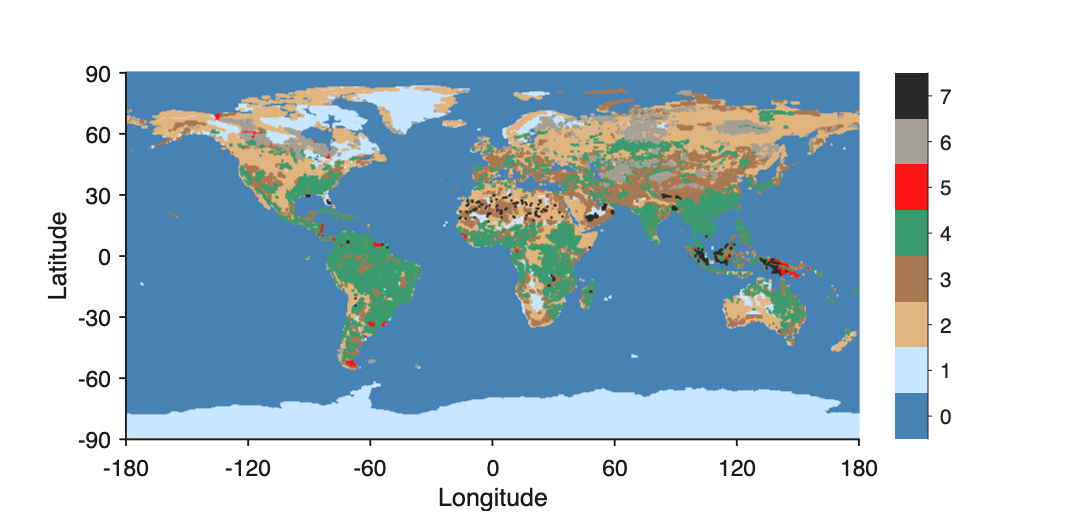}
\caption{Global soil type map from ERA5.}
\label{fig:ERA5_soil_type}
\end{figure}

Since ERA5 provides data on a fixed grid with $0.25^{\circ} \times 0.25^{\circ}$ resolution, each IASI observation is assigned the soil type of the nearest grid point in the $L^1$ norm (based on latitude and longitude differences) with respect to the coordinates of the IASI FoV center. The global spatial distribution of soil types used in this study is shown in Figure~\ref{fig:ERA5_soil_type}.

\noindent Table~\ref{tab:number_training_test} shows the total number of IASI observations (i.e., single spectral acquisitions, each corresponding to one FOV) contained in the training and test sets, divided by the periods of Spring, Summer, Autumn, Winter, and by soil type. 
\begin{table}[H]
\centering
\resizebox{\textwidth}{!}{%
\begin{tabular}{|c|cc|cc|cc|cc|}
\hline
& \multicolumn{2}{c|}{\textbf{Spring}} 
& \multicolumn{2}{c|}{\textbf{Summer}} 
& \multicolumn{2}{c|}{\textbf{Autumn}} 
& \multicolumn{2}{c|}{\textbf{Winter}} \\
\hline
\textbf{Soil Type} 
& \textbf{Train} & \textbf{Test} 
& \textbf{Train} & \textbf{Test} 
& \textbf{Train} & \textbf{Test} 
& \textbf{Train} & \textbf{Test} \\
\hline
0 & 157928 & 74600 & 158344 & 73560 & 158400 & 77184 & 152840 & 76720 \\
\hline
1 & 32080 & 16808 & 32016 & 16316 & 32620 & 16304 & 31216 & 15828 \\
\hline
2 & 25004 & 11184 & 24836 & 12676 & 24398 & 11816 & 24388 & 11916 \\
\hline
3 & 8100 & 5560 & 8336 & 4228 & 8516 & 4716 & 8000 & 5204 \\
\hline
4 & 8060 & 4972 & 8172 & 3428 & 7948 & 4152 & 7040 & 5088 \\
\hline
5 & 132 & 56 & 108 & 68 & 124 & 28 & 160 & 104 \\
\hline
6 & 1972 & 1292 & 1964 & 896 & 1932 & 1020 & 1896 & 880 \\
\hline
7 & 152 & 96 & 144 & 80 & 184 & 84 & 176 & 116 \\
\hline\hline
\textbf{TOTAL} 
& \textbf{233428} &  \textbf{114568}
& \textbf{232920} & \textbf{111252} 
& \textbf{234122} & \textbf{115304}
& \textbf{225716} & \textbf{115856} \\
\hline
\end{tabular}%
}
\caption{Total number of IASI observations (FOV), each associated with one full Level 1C spectrum, in the training and test sets for each season, divided by soil type.}
\label{tab:number_training_test}
\end{table}

\noindent To study the impact of geolocation on the algorithm, we explore two dataset configurations with respect to climatic zoning:

\begin{itemize}
    \item[(a)]\label{item:no_subdivision} No subdivisions: treating all data as one globally homogeneous set;
    \item[(b)]\label{item:climate_zones} Climate zones: splitting the data into five latitudinal bands representing typical climatic regions: Tropical, Temperate North/South, and Polar North/South.
\end{itemize}

The latitude boundaries that define each subdivision are summarized in Table~\ref{tab:subdivisions}.
\begin{table}[H]
\centering
\resizebox{\textwidth}{!}{ 
\begin{tabular}{|c|c|c|c|}
\hline
\textbf{Subdivision} & \textbf{Zone ID} & \textbf{Climate Zones} & \textbf{Latitude Range} \\
\hline
\hline
No subdivisions & - & All & [-90, 90] \\
\hline
\multirow{5}{*}{Climate zones} 
& 1 & Tropical & [-23, 23] \\
& 2 & Temperate North & [23, 66] \\
& 3 & Temperate South & [-66, -23] \\
& 4 & Polar North & [66, 90] \\
& 5 & Polar South & [-90, -66] \\
\hline
\end{tabular}
}
\caption{Latitude bounds defining the geographic subdivisions used in the preparation of training and test sets for cloud detection. The table includes two different geographic subdivisions: global (no subdivision) and five broad climate zones for spatially aware analysis. The Zone ID column provides a numerical reference used internally for data processing.}
\label{tab:subdivisions}
\end{table}

\noindent The datasets used in this study are structured along three dimensions of variability: season, surface type, and climate (also called geographic) zone, the latter including a configuration with no climatic subdivision, treating the data as a single global set. These dimensions are combined systematically, and the algorithm's performance is always reported as global accuracy, i.e., the overall percentage of correctly classified points aggregated over the full test set for each season and configuration.

\section{Application of the algorithm to IASI measurements}\label{sec:svm_to_IASI}
To apply the SVM framework introduced in Subsection~S2.1 to IASI Level~1C spectra, we adopt the clear and cloudy labeling defined in Section~\ref{sec:training_test_sets} and train binary classifiers on the corresponding hyperspectral inputs. For simplicity, we refer to the resulting implementation as the Cloud Identification SVM (CISVM).

Given the strong correlation and spectral oversampling of IASI radiances discussed in Section~\ref{sec:training_test_sets} and in Section~S1, the full spectrum is not used directly as SVM input. Instead, two complementary input representations are systematically compared: fixed-channel selection and Principal Component Analysis (PCA), the latter retaining a large fraction of the variance captured in the training set. As in the case of radiances, brightness temperatures (BT) can be used as an alternative input representation (see Subsection~S2.2).

A preliminary identification test using radiances was conducted to determine the fixed-channel baseline. Several configurations were compared, and the 45-channel subset proposed by Whitburn et al.~\cite{Whitburn_cloud_mask} achieved the highest classification accuracy (76.74\%); it is therefore adopted as the fixed-channel baseline in all subsequent evaluations (see Subsection~S2.3).

Combining this channel-selection strategy with the radiometric input choice (RAD or BT) and the climatic subdivision introduced in Section~\ref{sec:training_test_sets} (GLOBE or ZONES), eight distinct configurations are defined from the following three binary options:
\begin{itemize}
\item soil type subdivision only (GLOBE) versus soil type combined with climatic zones (ZONES),
\item fixed channel selection (CHANNELS) versus principal component analysis (PCA),
\item radiance input (RAD) versus brightness temperature input (BT).
\end{itemize}
The joint stratification (ZONES) inherently generates theoretically invalid combinations (e.g., tropical organic soils in polar zones), which correspond to empty data subsets. In addition, combinations yielding mono-class training sets (i.e., exclusively cloudy or exclusively clear profiles) are conservatively excluded to avoid biased decision boundaries, accounting for the slightly reduced sample size in the ZONES configurations.

\subsection{PCA and SVM Configuration}\label{subsubsec:PCA_SVM_configuration}
For the PCA-based configurations, dimensionality reduction was performed independently for each geographical subdivision and season, following a standardized pipeline with a cumulative explained variance threshold of 99.5\%. Across all configurations, the median percentage of retained principal components ranges between 1.41\% and 4.11\% of the original spectral dimensionality, confirming the strong redundancy of spectral infrared observations; the complete methodology and seasonal statistics are reported in Subsection~S2.5.

All final CISVM configurations employ a Gaussian radial basis function (RBF) kernel, with the Box Constraint fixed to $C=1$ and the kernel scale automatically estimated from the statistical properties of the training dataset. For the selected best-performing configuration (RAD-GLOBE-PCA), the resulting kernel scale is $\mathrm{KernelScale} \approx 64.87$; the complete formulation and the statistical characterization across all spectral representations are reported in Subsection~S2.6.
\subsection{Results}\label{subsec:results}
\begin{table}[H]
\centering
\resizebox{\textwidth}{!}{
\begin{tabular}{|c|c|c||c|c|c|c|c|}
\hline
\textbf{BT or RAD} & 
\makecell{\textbf{Geography}\\\textbf{Division}} & 
\makecell{\textbf{Spectral}\\\textbf{Domain}} & 
\textbf{SPRING} & 
\textbf{SUMMER} & 
\textbf{AUTUMN} & 
\textbf{WINTER} & 
\makecell{\textbf{\%}\\\textbf{TOTAL}} \\
\hline
\hline

BT & GLOBE & CHANNELS &
\begin{tabular}{c}
98896/114568 \\
86.32\%
\end{tabular} &
\begin{tabular}{c}
97468/111252 \\
87.61\%
\end{tabular} &
\begin{tabular}{c}
99740/115304 \\
86.50\%
\end{tabular} &
\begin{tabular}{c}
100808/115856 \\
87.01\%
\end{tabular} &
\begin{tabular}{c}
396912/456980 \\
86.86\%
\end{tabular} \\
\hline

BT & ZONES & CHANNELS &
\begin{tabular}{c}
97932/114568 \\
85.48\%
\end{tabular} &
\begin{tabular}{c}
96660/111252 \\
86.88\%
\end{tabular} &
\begin{tabular}{c}
99024/115300 \\
85.88\%
\end{tabular} &
\begin{tabular}{c}
100108/115856 \\
86.41\%
\end{tabular} &
\begin{tabular}{c}
393724/456976 \\
86.16\%
\end{tabular} \\
\hline

BT & GLOBE & PCA &
\begin{tabular}{c}
100348/114568 \\
87.59\%
\end{tabular} &
\begin{tabular}{c}
99076/111252 \\
89.06\%
\end{tabular} &
\begin{tabular}{c}
101716/115304 \\
88.22\%
\end{tabular} &
\begin{tabular}{c}
102708/115856 \\
88.65\%
\end{tabular} &
\begin{tabular}{c}
403848/456980 \\
88.37\%
\end{tabular} \\
\hline

BT & ZONES & PCA &
\begin{tabular}{c}
99236/114568 \\
86.62\%
\end{tabular} &
\begin{tabular}{c}
98660/111252 \\
88.68\%
\end{tabular} &
\begin{tabular}{c}
100800/115300 \\
87.42\%
\end{tabular} &
\begin{tabular}{c}
102012/115856 \\
88.05\%
\end{tabular} &
\begin{tabular}{c}
400708/456976 \\
87.69\%
\end{tabular} \\
\hline

RAD & GLOBE & CHANNELS &
\begin{tabular}{c}
98692/114568 \\
86.14\%
\end{tabular} &
\begin{tabular}{c}
96772/111252 \\
86.98\%
\end{tabular} &
\begin{tabular}{c}
98844/115304 \\
85.72\%
\end{tabular} &
\begin{tabular}{c}
99936/115856 \\
86.26\%
\end{tabular} &
\begin{tabular}{c}
394244/456980 \\
86.27\%
\end{tabular} \\
\hline

RAD & ZONES & CHANNELS &
\begin{tabular}{c}
97980/114568 \\
85.52\%
\end{tabular} &
\begin{tabular}{c}
96732/111252 \\
86.95\%
\end{tabular} &
\begin{tabular}{c}
98976/115300 \\
85.84\%
\end{tabular} &
\begin{tabular}{c}
99908/115856 \\
86.23\%
\end{tabular} &
\begin{tabular}{c}
393596/456976 \\
86.13\%
\end{tabular} \\
\hline

RAD & GLOBE & PCA &
\begin{tabular}{c}
100164/114568 \\
87.43\%
\end{tabular} &
\begin{tabular}{c}
99744/111252 \\
89.66\%
\end{tabular} &
\begin{tabular}{c}
101932/115304 \\
88.40\%
\end{tabular} &
\begin{tabular}{c}
102664/115856 \\
88.61\%
\end{tabular} &
\begin{tabular}{c}
\textbf{404504/456980} \\
\textbf{88.52}\%
\end{tabular} \\
\hline

RAD & ZONES & PCA &
\begin{tabular}{c}
99412/114568 \\
86.77\%
\end{tabular} &
\begin{tabular}{c}
98620/111252 \\
88.65\%
\end{tabular} &
\begin{tabular}{c}
101056/115300 \\
87.65\%
\end{tabular} &
\begin{tabular}{c}
102280/115856 \\
88.28\%
\end{tabular} &
\begin{tabular}{c}
401368/456976 \\
87.83\%
\end{tabular} \\
\hline

\end{tabular}
}
\caption{Cloud detection accuracy results using SVM for different physical quantities BT or RAD, for globe or climate zones divisions, and spectral input types for CHANNELS or PCA. Results are reported for all seasons together with the overall classification accuracy.}
\label{tab:svm_results_summary}
\end{table}

\noindent The results are shown in Table~\ref{tab:svm_results_summary}. To assess the statistical significance of the differences between test outcomes, we applied the z-test (see e.g. \cite{moore2013}) to pairs of tests. The corresponding significance levels are summarized in Table~\ref{tab:significance_matrix}, where tests are identified by the initials of the first three columns of Table~\ref{tab:svm_results_summary} and listed in decreasing accuracy order. Statistical significance is indicated as follows: ns (not significant), * ($p<0.05$, significant), ** ($p<0.01$, highly significant), and *** ($p<0.001$, very highly significant).

\begin{table}[H]
\centering
\begin{tabular}{lcccccccc}
\hline
 & RGP & BGP & RZP & BZP & BGC & RGC & BZC & RZC \\
\hline
RGP &  & ns & *** & *** & *** & *** & *** & *** \\
BGP &  &  & *** & *** & *** & *** & *** & *** \\
RZP &  &  &  & ns & *** & *** & *** & *** \\
BZP &  &  &  &  & *** & *** & *** & *** \\
BGC &  &  &  &  &  & *** & *** & *** \\
RGC &  &  &  &  &  &  & ns & * \\
BZC &  &  &  &  &  &  &  & ns \\
RZC &  &  &  &  &  &  &  &  \\
\hline
\end{tabular}
\caption{Significance of z-tests between test pairs: ns (not significant), * ($p<0.05$), ** ($p<0.01$), *** ($p<0.001$).}
\label{tab:significance_matrix}
\end{table}

\noindent Due to the large sample size, even minor performance differences are statistically significant; therefore, the choice of configuration must be guided by the actual magnitude of the accuracy improvement. While the choice between radiance and brightness temperature yields a negligible impact, adopting PCA provides a highly significant and practically relevant accuracy gain of 1–2 percentage points over explicit channel selection. This improvement occurs because cloud presence affects the entire infrared spectrum: retaining broader spectral variance via PCA enhances class separability, with minimal computational overhead during inference, see Subsection~\ref{subsec:computational_cost_and_memory}. Consequently, we select the overall highest-accuracy configuration for the subsequent comparison against independent cloud cover data.

\subsection{Surface-Type Dependence of Classification Performance}\label{subsec:surface_classification_performance}
The RAD-GLOBE-PCA configuration, reported in Table~\ref{tab:svm_results_summary}, achieved the highest overall classification accuracy and was therefore selected for further analysis. The following analysis examines its performance across different surface types, with the aim of identifying the physical factors that drive the observed variability in cloud detection performance. The classification metrics used throughout this analysis (Accuracy, Cloud/Clear Recall, Precision, Balanced Accuracy, and F1-score) are defined in Table~S6.

\begin{figure}[H]
\centering
\includegraphics[width=1\textwidth]{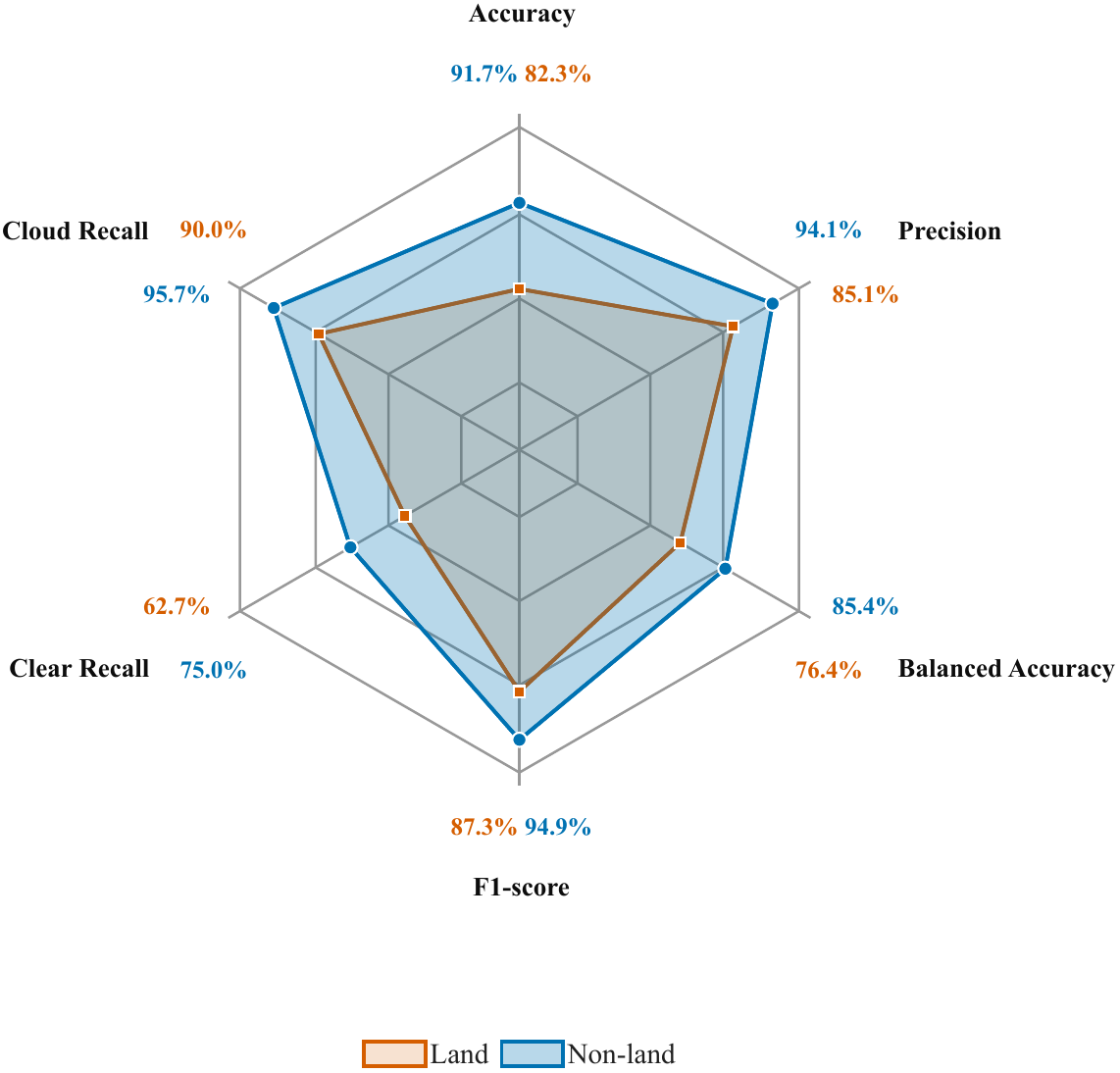}
\caption{Comparison of cloud detection performance over land and non-land surfaces for the selected RAD-GLOBE-PCA configuration, with metrics averaged across the four seasons and weighted by the number of observations in each season and soil type. All metrics are expressed as percentages. The F1-score, originally defined in $[0,1]$, is here multiplied by 100 for consistency of presentation.}
\label{fig:land_nonland_analysis}
\end{figure}

Figure~\ref{fig:land_nonland_analysis} summarizes the performance of the selected configuration over land and non-land surfaces, averaged across the four seasons and weighted by the number of observations in each season and soil type. 

As illustrated in Figure~\ref{fig:land_nonland_analysis}, the classification performance exhibits a systematic degradation over continental surfaces. While the F1-score and Cloud Recall reflect a generally robust configuration, the most pronounced discrepancies occur in Accuracy and Clear Recall. In terms of classification dynamics, this indicates that over land, the model is prone to false cloud alarms, frequently misclassifying true clear-sky pixels as cloudy.

This behaviour is driven by the spectral and radiative heterogeneity of continental surfaces. Variations in surface emissivity, skin temperature, and vegetation cover broaden the natural variance of clear sky radiances in the thermal infrared channels; modelling these distinct emissivity profiles remains a well-known challenge in spectral data retrieval~\cite{sgheri2024}. Specifically, surface types such as cold, snow-covered, or high-altitude terrain exhibit spectral and radiometric signatures that partially overlap with those of clouds, particularly in terms of brightness temperature and thermal contrast, thereby narrowing the classifier's decision margin. This overlap causes a substantial fraction of genuinely clear-sky scenes to be misclassified as cloudy, which directly explains the observed drop in Clear Recall. The inverse pattern is comparatively rare: Cloud Recall remains consistently high across both land and non-land surfaces, indicating that clouds are correctly identified regardless of the underlying surface type. This asymmetry suggests that the classification error is predominantly one-directional, with clear-sky scenes being misidentified as cloudy far more frequently than the reverse.

The persistently high and stable Cloud Recall observed across both domains confirms the configuration's sensitivity to genuinely cloudy scenes, even against complex continental backgrounds. To better understand the specific surface properties driving the clear-sky misclassifications, the classification performance is further investigated in the following soil-type analysis.

\begin{figure}[H]
\centering
\includegraphics[width=1\textwidth]{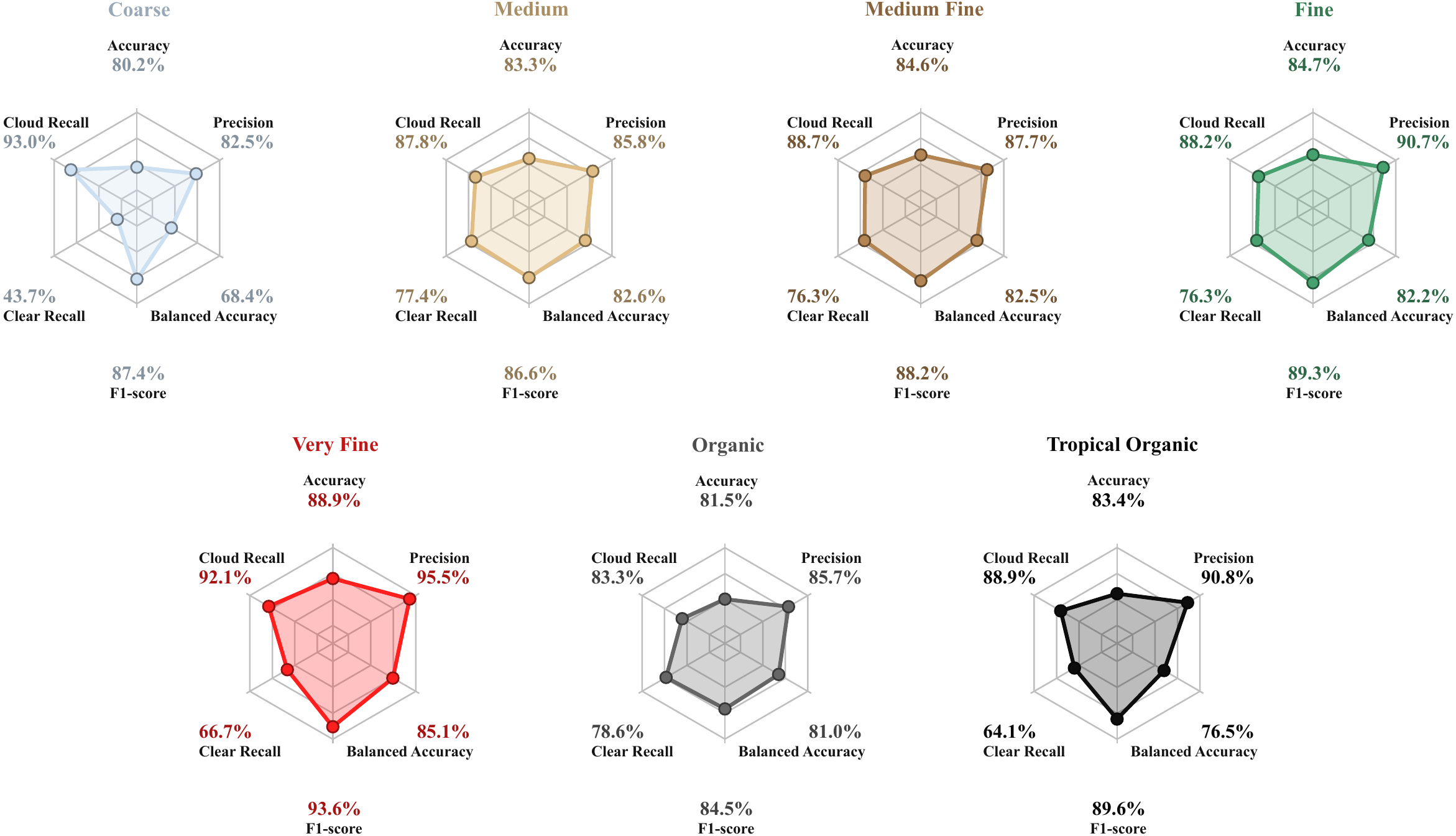}
\caption{Cloud detection performance of the RAD-GLOBE-PCA configuration 
stratified by soil type. Each radar chart reports seasonal-mean values; 
all metrics are expressed as percentages.}
\label{fig:soil_type_analysis}
\end{figure}

Figure~\ref{fig:soil_type_analysis} decomposes the land-surface classification results by soil type. Cloud Recall is stable across all classes, ranging from $83.3\%$ (Organic) to $93.0\%$ (Coarse), while Clear Recall spans approximately 35 percentage points, confirming it as the primary driver of inter-class variability.

Coarse soils yield the lowest Clear Recall ($43.7\%$) despite the highest Cloud Recall ($93.0\%$). Dominant over glacial and periglacial surfaces, these regions exhibit low brightness temperature contrast between snow and ice-covered surfaces and overlying cloud tops in the thermal infrared. This spectral similarity causes clear-sky observations to project into cloudy regions of the spectral feature space, driving a high rate of false cloud detections.

This seasonal dependency was further quantified by aggregating classification errors into three latitudinal bands (North: [60\textdegree,90\textdegree] -- Mid: [-60\textdegree,60\textdegree] -- South: [-90\textdegree,-60\textdegree]), consistent with the geographic distribution of Coarse soils over glacial and periglacial surfaces. As reported in Table~\ref{tab:seasonal_variability}, the polar bands exhibit a standard deviation approximately six to seven times larger than that of the mid-latitude band (14.42\% and 11.54\%, respectively, versus 1.98\%), with corresponding seasonal error ranges of 34.7\% and 28.0\%, compared to only 4.6\% for the Mid band. This disparity indicates that classification error over polar surfaces is strongly modulated by seasonal forcing rather than reflecting a fixed geographic bias intrinsic to the classifier.

\begin{table}[H]
\centering
\resizebox{\textwidth}{!}{
\begin{tabular}{|c|c|c|c|}
\hline
\textbf{Latitudinal band} & \textbf{Mean error (\%)} & \textbf{Std. dev. (\%)} & \textbf{Range (\%)} \\
\hline
\hline
North [60\textdegree, 90\textdegree]   & 21.32 & 14.42 & 34.68 \\
\hline
Mid [-60\textdegree, 60\textdegree]    & 18.29 & 1.98  & 4.59  \\
\hline
South [-90\textdegree, -60\textdegree] & 19.73 & 11.54 & 28.01 \\
\hline
\end{tabular}
}
\caption{Seasonal variability of classification error across latitudinal 
bands, restricted to observations classified as Coarse soil type. Mean and standard deviation are computed across the four seasons (Winter, Spring, Summer, Autumn).}
\label{tab:seasonal_variability}
\end{table}

\begin{figure}[H]
\centering
\includegraphics[width=0.7\textwidth]{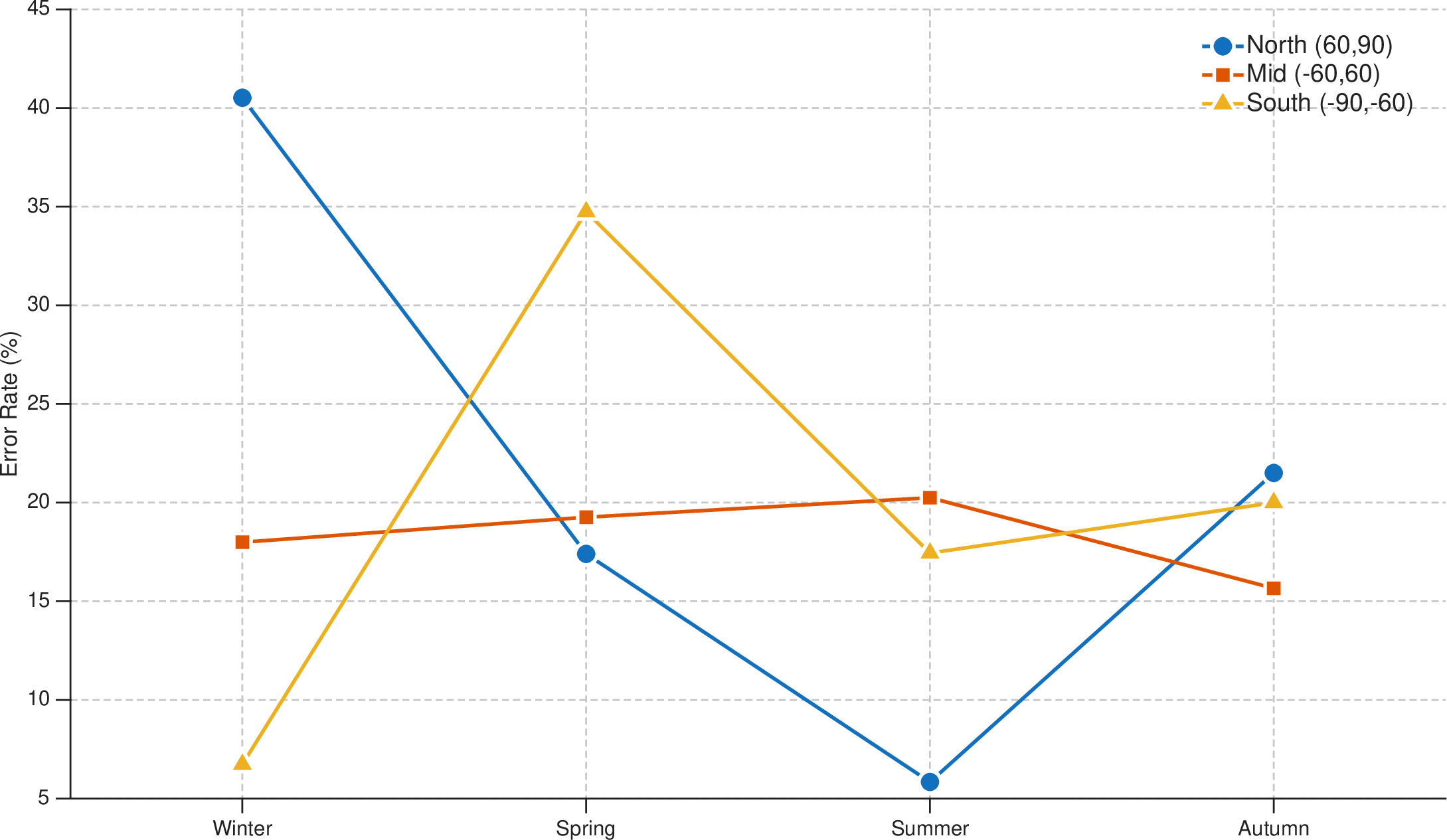}
\caption{Seasonal variation of classification error across latitudinal 
bands, restricted to observations classified as Coarse soil type.}
\label{fig:seasonal_bands}
\end{figure}

\noindent Figure~\ref{fig:seasonal_bands} illustrates this behaviour: the North and South bands display pronounced seasonal oscillations, peaking during local Winter and Spring, respectively, whereas the Mid-latitude band remains nearly stationary throughout the annual cycle. The statistical significance of this association (\(\chi^2 = 2298.6\), \(df=6\), \(p<0.001\)) confirms that the spatial pattern of clear-sky misclassification is tightly coupled to the seasonal extent of snow and ice cover, rather than representing a static limitation of the classifier itself.

\noindent Tropical Organic soils show a similarly reduced Clear Recall ($64.1\%$). In these regions, high precipitable water content reduces the brightness temperature contrast between cloud tops and the warm, moist surface, generating substantial cluster overlap in the spectral feature space.

Medium, Medium Fine, and Fine soils, widespread across temperate zones, exhibit the most consistent performance, with Accuracy above $83\%$ and Balanced Accuracy around $82\%$. These results suggest a better spectral separability between clear sky and cloudy observations for these surface types, possibly related to more stable surface emissivity characteristics and stronger radiative separation between surface and cloud signatures.

Organic soils, associated with boreal and sub-arctic regions, achieve the most symmetric performance (Cloud Recall $83.3\%$, Clear Recall $78.6\%$, Balanced Accuracy $81.0\%$). The reduced confusion may be related to the relatively stable radiative properties of these environments, which limit the overlap between clear and cloudy observations in the spectral feature space.

Very Fine soils yield the highest F1-score ($93.6\%$) and Precision ($95.5\%$). This strong performance can be attributed to their higher water retention capacity and consequently higher thermal inertia; this reduces short term surface temperature variability, ultimately contributing to more consistent and easily recognizable thermal infrared signatures.

Overall, the results confirm that classification uncertainty over land originates primarily from the discrimination of clear-sky observations, whose radiative signatures over thermally and spectrally heterogeneous surfaces overlap with those of cloudy scenes in the spectral feature space.

To quantify the complexity of the trained SVM, we analyze the Support Vector Ratio (SVRatio), which provides an empirical proxy for classification complexity; its formal definition and numerical details are reported in Subsection~S2.7.

\begin{figure}[H]
\centering
\includegraphics[width=1\textwidth]{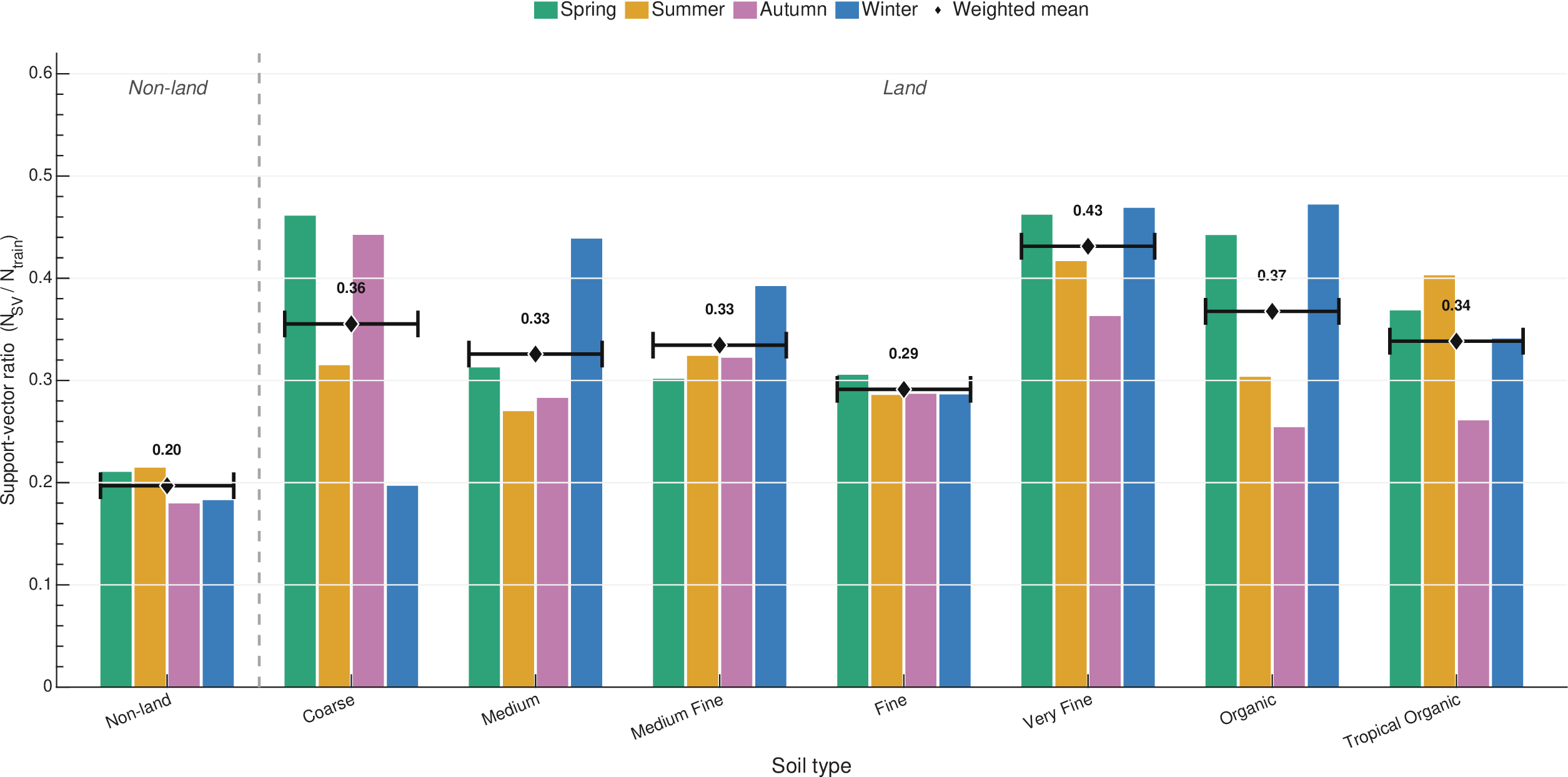}
\caption{Support Vector Ratio ($N_{\text{SV}}/N_{\text{train}}$) of the RAD-GLOBE-PCA configuration across surface types and seasons. Horizontal bars indicate the weighted mean for each surface class.}
\label{fig:soil_type_svratio}
\end{figure}

Figure~\ref{fig:soil_type_svratio} reveals that classifier complexity depends strongly on surface type: non-land surfaces exhibit the lowest SVRatio, whereas all land classes display substantially larger values, consistent with the surface spectral and thermal heterogeneity already discussed in this Subsection. Two regimes emerge: for some soil types high complexity coincides with degraded performance, whereas for others it coexists with excellent metrics, indicating that decision-boundary complexity and class separability are not equivalent.

The variability of SVRatio across soil classes substantially exceeds its seasonal variability, supporting the conclusion that land-surface heterogeneity is the primary driver of classifier complexity in infrared cloud detection.
\subsection{Computational Cost and Memory Footprint}\label{subsec:computational_cost_and_memory}

In addition to classification accuracy, the computational efficiency of the CISVM configurations was analyzed in order to evaluate the practical impact of the adopted spectral representations on both the training and operational phases of the algorithm. For each test configuration, the computational cost was evaluated separately for the SVM training phase and the prediction phase, and global statistics were computed using weighted averages to obtain representative per-sample execution times over the entire dataset; the formal definitions are provided in Subsection~S2.8.

\begin{table}[htbp]
\centering
\small
\begin{tabular}{|c|c|c|}
\hline
\textbf{Spectral} &
\makecell{\textbf{Total Training}\\\textbf{Time [min]}} &
\makecell{\textbf{Prediction}\\\textbf{Time/sample [ms]}} \\
\hline
CHANNELS & 100.86  & \makecell{2.72 \\ {[1.57, 5.29]}} \\
\hline
PCA      & 168.82  & \makecell{13.96 \\ {[6.09, 47.25]}} \\
\hline
\end{tabular}
\caption{Cumulative total training time and prediction cost per sample, for the CHANNELS and PCA configurations aggregated over all seasons, geographical subdivisions, and spectral input types. Due to the strongly right-skewed distribution of computational costs, prediction times are reported as the median followed by the interquartile range (first and third quartiles) in brackets. A detailed breakdown by soil type is provided in the Supplementary Material.}
\label{tab:computational_cost_summary}
\end{table}

Table~\ref{tab:computational_cost_summary} shows that PCA-based configurations require a cumulative total training time of approximately 169~min, compared to about 101~min for the CHANNELS configurations, and a substantially larger median prediction cost (13.96~ms per spectrum versus 2.72~ms), corresponding to an approximately fivefold increase. The moderate overhead observed in the total training time indicates that the dimensionality reduction procedure does not critically affect the overall training efficiency of the SVM classifier, whereas the prediction phase is considerably more demanding for PCA configurations.

This behavior is primarily associated with the large effective dimensionality retained by the PCA representation, as summarized in Table~S4, where the median number of retained principal components ranges from roughly 120 to 350 depending on the season and configuration. The resulting high-dimensional feature spaces increase the cost of evaluating the Gaussian kernel during classification, and the kernel parametrization adopted in this work further couples the evaluation cost to the dimensionality and statistical properties of each training representation, leading to experiment-dependent prediction times for PCA, in contrast to the more stable and lower prediction cost observed for the fixed 45 channel input.

Nevertheless, the PCA-based configurations simultaneously provide the highest classification accuracies and the most stable kernel-scale statistics among all tested configurations, highlighting a clear trade-off between classification performance and computational efficiency: PCA improves the separability and regularity of the spectral feature space at the expense of an increased operational computational cost, particularly in the prediction phase.

The memory footprint of the deployed model was evaluated by isolating the components strictly required for inference on new spectra: the trained SVM object, the truncated PCA transformation matrix, and the standardization parameters. Since the CISVM architecture loads a single soil-type-specific sub-model at a time, the relevant operational footprint corresponds to the largest individual sub-model, consistently associated with the Non-land class, and ranges from approximately 170~MB in Summer to about 477~MB in Spring (Table~S4). Even under a more conservative scenario where all eight soil-type sub-models are kept simultaneously in memory, the cumulative footprint remains modest, between 305~MB and 747~MB across seasons, confirming that the memory requirements of the deployed system are compatible with operational or onboard processing constraints.
\section{Comparison with MODIS cloud mask\label{sec:cloud_mask_and_MODIS_data}}
Having evaluated the CISVM against the IASI Level 1C reference, we now provide an external consistency assessment against an independent passive cloud product. Specifically, we compare the best performing CISVM configuration (RAD-GLOBE-PCA) and the IASI reference itself against MODIS data to verify whether their large scale spatial behaviour remains consistent across different observing systems.

In the following, the term \textit{cloud mask} refers to a binary classification that indicates the presence or absence of clouds, in which each grid point is associated with a single cloud observation per day, thus representing the cloud conditions at that location and scan time.

As shown in Table~\ref{tab:svm_results_summary}, the CISVM classification accuracy is remarkably similar across all four seasons. Given this consistency, we restrict the MODIS comparison to Winter and Summer, the two seasons with the largest climatological contrast, while remaining broadly representative of the other two. This choice keeps the presentation concise while remaining consistent with the four season assessment reported above.

For the independent comparison against MODIS, we consider one full day of IASI measurements for Winter and one for Summer, each composed of 15 consecutive orbits; the complete list of dates, times, and orbit ID is reported in Table~S7.

\noindent For comparison, we use data derived from the MODIS Level 2 cloud products~\cite{justice2002}. MODIS instruments are onboard two different satellites: Terra and Aqua. By combining the MODIS Terra (MOD06\_L2)~\cite{mod06_l2} and MODIS Aqua (MYD06\_L2)~\cite{myd06_l2} datasets, we obtain two daily observations for each geolocation, thereby increasing the number of potential coincidences with IASI measurements.

Both datasets provide cloud mask information at a spatial resolution of 1~km. MOD06\_L2 corresponds to Terra's morning overpass, while MYD06\_L2 refers to Aqua’s afternoon orbit, offering complementary temporal coverage.

We preliminarily test the CISVM algorithm on the newly selected IASI test set over the entire globe (i.e., prior to determining coincidences with the MODIS instrument), obtaining an accuracy of 87.13\% in Winter and 88.22\% in Summer, resulting in an overall accuracy of 87.67\%. These values are very close to those obtained in the previous test, confirming the robustness of the CISVM method.

When comparing cloud products, a strict temporal threshold is necessary because of the high spatial and temporal variability of clouds. In this study, collocation between MODIS and IASI measurements was based on both temporal coincidence and spatial overlap. We set the temporal coincidence threshold to 20 minutes, which represents a compromise between limiting cloud evolution effects and preserving a sufficiently large test set.

Spatial matching was performed by exploiting the IASI field of view (FoV) geometry. Since MODIS cloud products are provided on a 1~\si{km} grid, for each collocated IASI observation we identified all MODIS pixels falling within the corresponding IASI FoV and averaged their cloud fraction values to obtain a representative MODIS cloud fraction at the IASI spatial scale. Therefore, no additional spatial distance threshold between pixel centers was introduced, as the spatial collocation was directly defined by the footprint overlap.

For consistency, the same cloud fraction threshold of 0.1 adopted for the IASI L1 cloud cover product is also applied to the collocated MODIS cloud fraction, so that differences between masks reflect sensor and retrieval behaviour rather than a change in the binary decision criterion. 

Although all three satellites operate in sun-synchronous orbits, they fly at different nominal altitudes. This difference in altitude strictly determines their different orbital periods and the resulting longitudinal spacing between successive orbits. As a consequence, the number of temporal and spatial coincidences is highly variable and depends on the selected day. The test days were therefore chosen among those maximizing the number of coincidences.

Coincidences between the IASI and MODIS orbits can be visually inspected using the orbit tracks and timing features available on the NASA Worldview website~\cite{nasa_worldview2025} for MODIS Aqua, MODIS Terra, and MetOp-C.

Due to the different ascending node times, most coincidences along the orbit occur with the MODIS Terra satellite. MODIS Aqua contributes only a limited number of coincidences, primarily in the polar regions, which are sampled at every orbit.

With this configuration, we obtain 1529 coincidences for the Winter test set and 2949 for the Summer test set.

Thus, we obtain the following three classifications on the same test set:
\begin{itemize}
    \item \textbf{IASI-THR}: binary classification obtained by applying the threshold 0.1 using the IASI L1 cloud fraction data;
    \item \textbf{IASI-CISVM}: binary classification obtained using the CISVM algorithm trained on IASI L1 cloud fraction data and the same threshold;
    \item \textbf{MODIS-THR}: binary classification obtained by applying the same threshold 0.1 to MODIS cloud fraction data obtained from the MOD06\_L2 and MYD06\_L2 products.
\end{itemize}

Table~\ref{tab:accuracy_results} reports the percentage agreement for each pairwise comparison of these classifications, for both Winter and Summer.

\begin{table}[H]
\centering
\begin{tabular}{|l|c|c|}
\hline
\textbf{Dataset comparison} & \textbf{Winter} & \textbf{Summer} \\
\hline\hline
IASI-THR vs MODIS-THR  & 76.85\% & 79.89\% \\
IASI-THR vs IASI-CISVM   & 88.62\% & 89.01\% \\
IASI-CISVM vs MODIS-THR  & 73.84\% & 80.91\% \\
\hline
\end{tabular}
\caption{Percentage agreement between the binary cloud classifications from the IASI-THR, MODIS-THR, and IASI-CISVM datasets for Winter and Summer.}\label{tab:accuracy_results}
\end{table}

\noindent The three pairwise comparisons in Table~\ref{tab:accuracy_results} quantify complementary aspects of the problem. The IASI-THR vs MODIS-THR agreement measures the structural discrepancy between two independent passive cloud products, the IASI-THR vs IASI-CISVM agreement measures how faithfully the classifier reproduces the operational IASI reference on the collocated subset, and the IASI-CISVM vs MODIS-THR agreement measures the external consistency of the learned infrared-only classification with respect to an independent sensor. Interpreting these values jointly is essential, because any disagreement between CISVM and MODIS combines both inter-sensor differences and the residual classification error inherited from training on the AVHRR-derived IASI reference. 

\begin{table}[H]
\centering
\footnotesize
\setlength{\tabcolsep}{4pt}
\renewcommand{\arraystretch}{1.2}

\begin{tabular}{
    >{\centering\arraybackslash}p{2cm}
    >{\centering\arraybackslash}p{2cm}
    >{\centering\arraybackslash}p{2cm}
    >{\centering\arraybackslash}p{2cm}
    >{\centering\arraybackslash}p{2cm}
    >{\centering\arraybackslash}p{2cm}
}
\multicolumn{3}{c}{\textbf{Confusion Matrices – Winter}} &
\multicolumn{3}{c}{\textbf{Confusion Matrices – Summer}} \\

$\begin{bmatrix}
173 & 129 \\
225 & 1002
\end{bmatrix}$ &

$\begin{bmatrix}
192 & 110 \\
64 & 1163
\end{bmatrix}$ &

$\begin{bmatrix}
127 & 129 \\
271 & 1002
\end{bmatrix}$ &

$\begin{bmatrix}
276 & 200 \\
393 & 2080
\end{bmatrix}$ &

$\begin{bmatrix}
250 & 226 \\
98 & 2375
\end{bmatrix}$ &

$\begin{bmatrix}
227 & 121 \\
442 & 2159
\end{bmatrix}$ \\
\end{tabular}

\caption{Confusion matrices for Winter (left block) and Summer (right block). For each block: IASI-THR vs. MODIS-THR (left), IASI-THR vs. IASI-CISVM (center), and IASI-CISVM vs. MODIS-THR (right).}
\label{tab:conf_bmatrix}
\end{table}

\noindent The confusion matrices in Table~\ref{tab:conf_bmatrix} for Winter and Summer show the level of agreement between pairs of classifiers: IASI-THR, MODIS-THR, and IASI-CISVM. Each matrix compares the outputs of two classifiers on the same dataset.The rows correspond to the predictions of the first classifier in the pair, while the columns correspond to those of the second classifier.

For each ordered pair (as specified in the table caption), element (1,1) indicates the number of points classified as \emph{clear} by both classifiers, while element (2,2) indicates agreement on the \emph{cloudy} class. Element (1,2) counts the number of points classified as \emph{clear} by the first classifier and \emph{cloudy} by the second, whereas element (2,1) represents the reverse situation.

The confusion matrices further show that the disagreement is not symmetric across classes. As illustrated in detail by the spatial analysis below, this discrepancy is strongly conditioned by the underlying surface regime. In particular, a substantial fraction of the mismatches corresponds to scenes identified as cloudy by IASI-based products and clear by MODIS, especially under polar conditions, which is consistent with the reduced cloud surface radiative contrast discussed in Section~\ref{sec:svm_to_IASI}. This behaviour indicates that the disagreement is driven less by random classification noise than by physically structured observing differences across challenging radiative environments.

\begin{figure}[htbp]
    \centering
    \includegraphics[width=\textwidth, trim=20 110 20 60, clip]{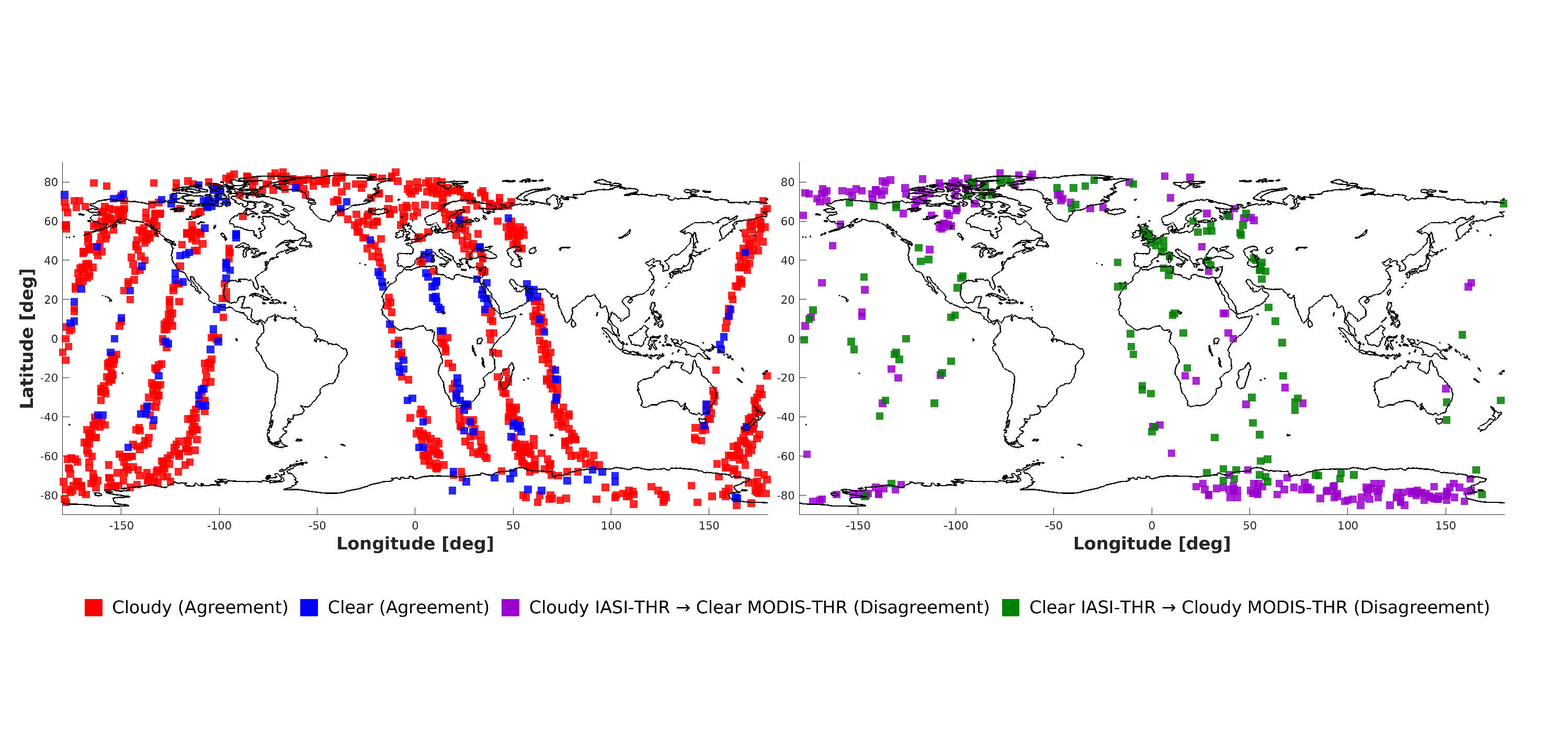}
    \includegraphics[width=\textwidth, trim=20 110 20 60, clip]{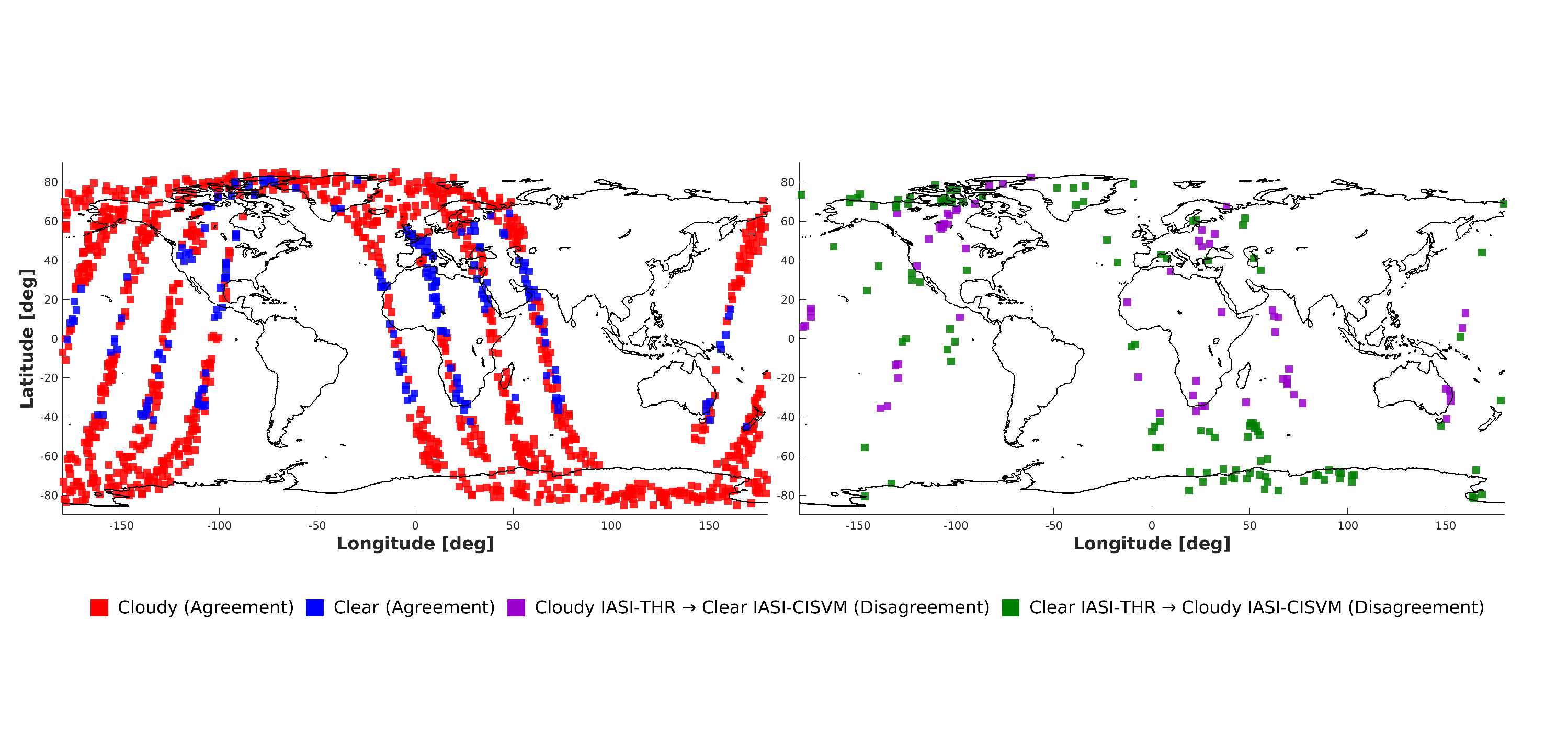}
    \includegraphics[width=\textwidth, trim=20 0 20 60, clip]{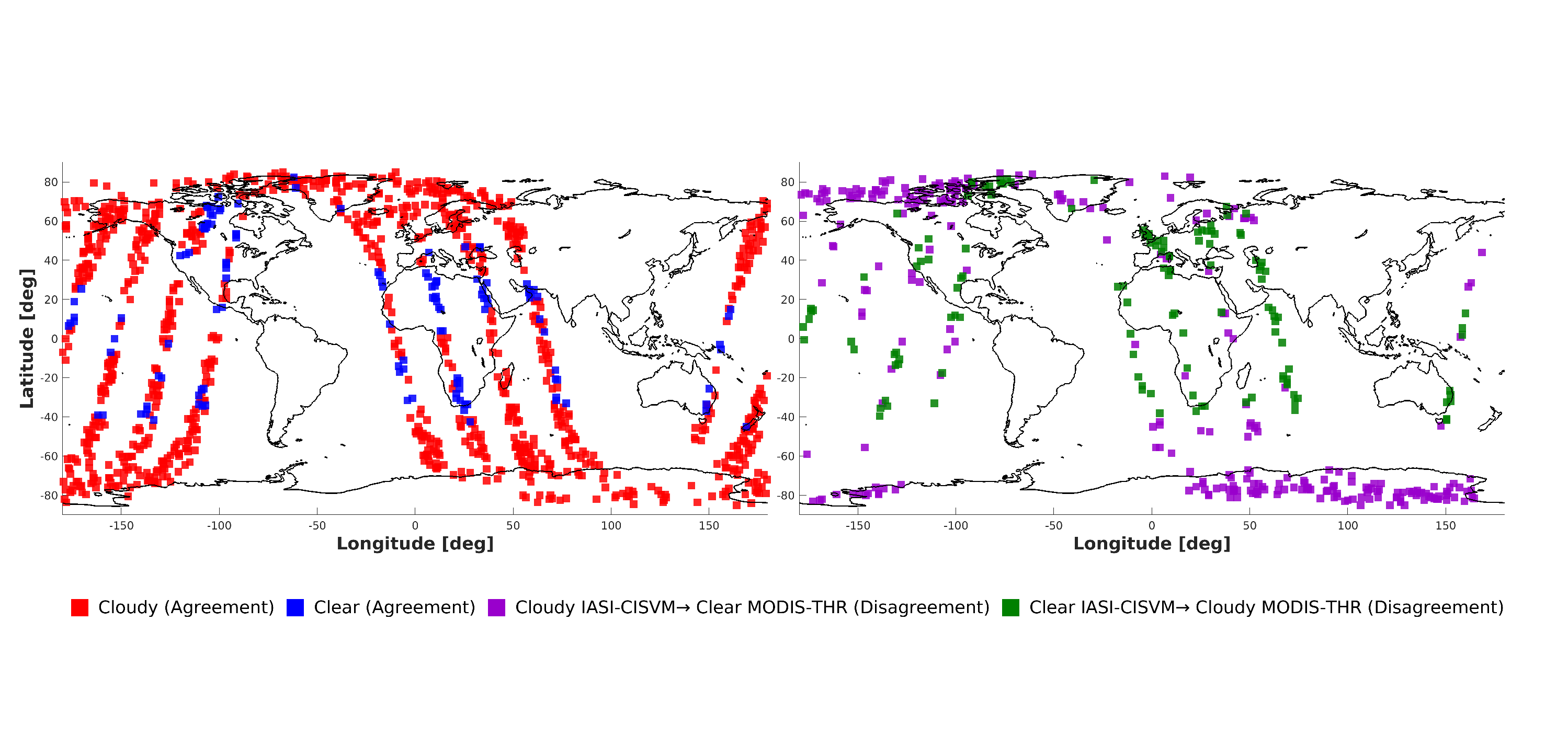}
    \caption{Differences in cloud detection agreement between the three cloud masks, at locations and times where all are available, for Winter. The figures in the panels are: IASI-THR vs MODIS-THR (top), IASI-THR vs IASI-CISVM (center), and IASI-CISVM vs MODIS-THR (bottom).}
    \label{fig:results_modis_svm_jan}
\end{figure}

\begin{figure}[htbp]
    \centering
    \includegraphics[width=\textwidth, trim=20 110 20 60, clip]{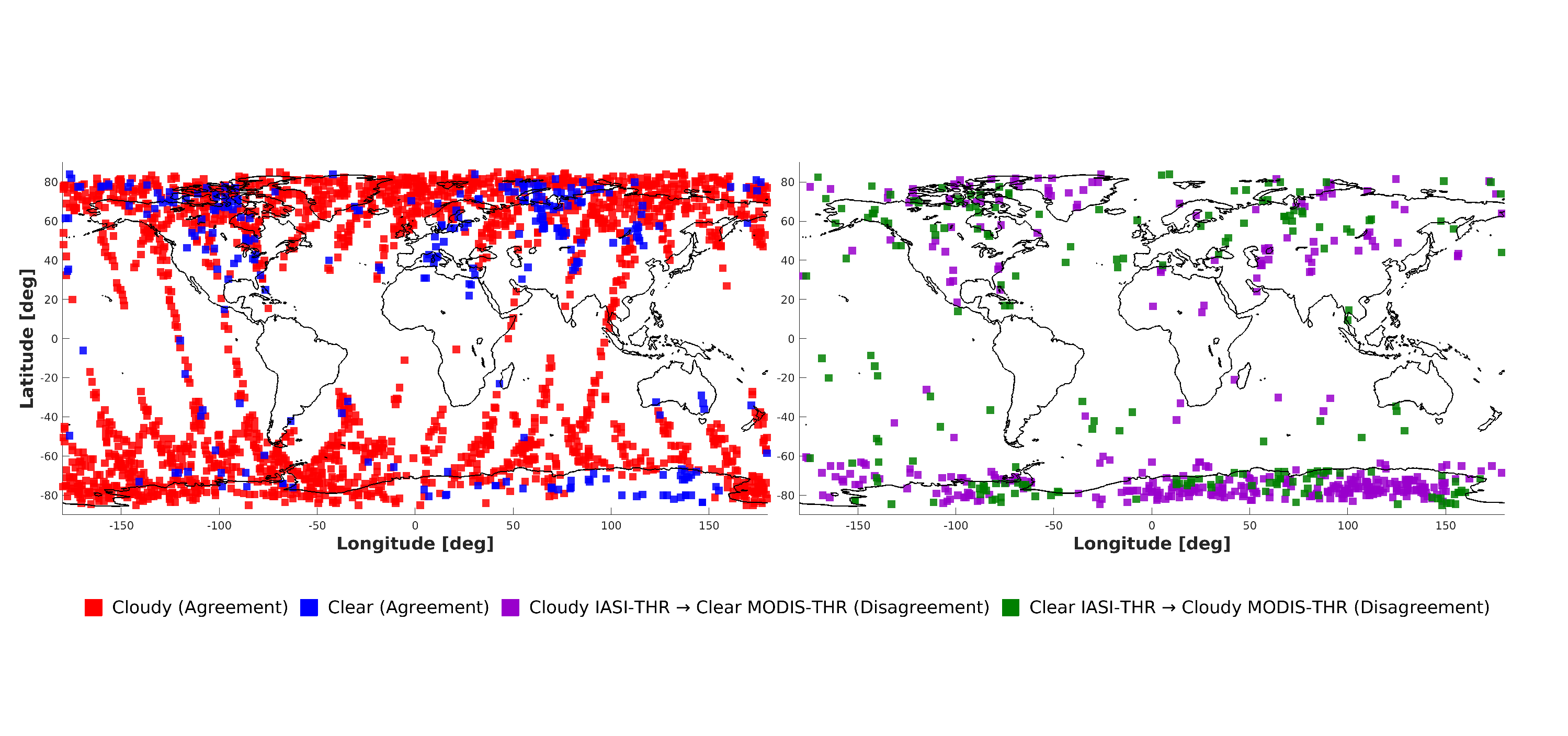}
    \includegraphics[width=\textwidth, trim=20 110 20 60, clip]{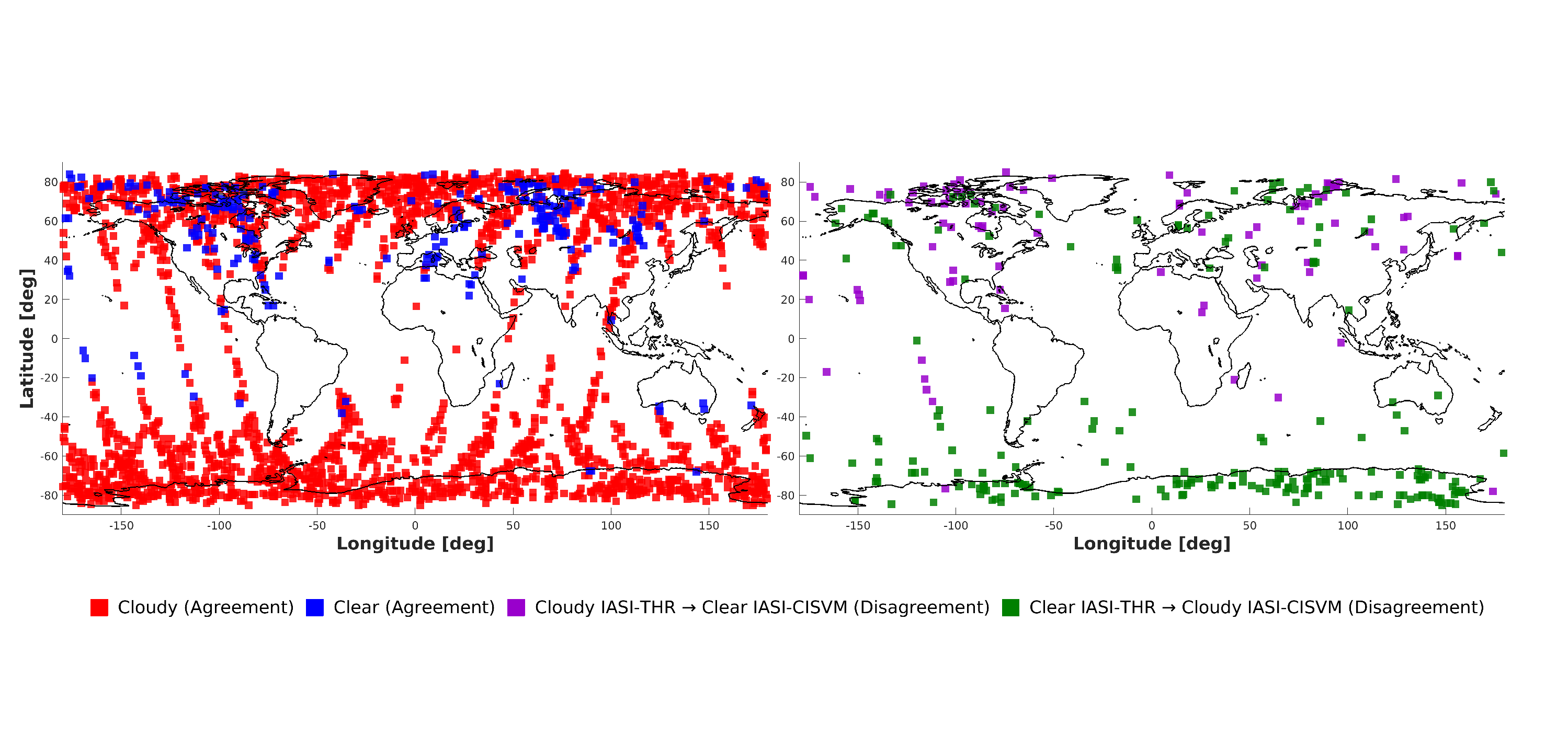}
    \includegraphics[width=\textwidth, trim=20 0 20 60, clip]{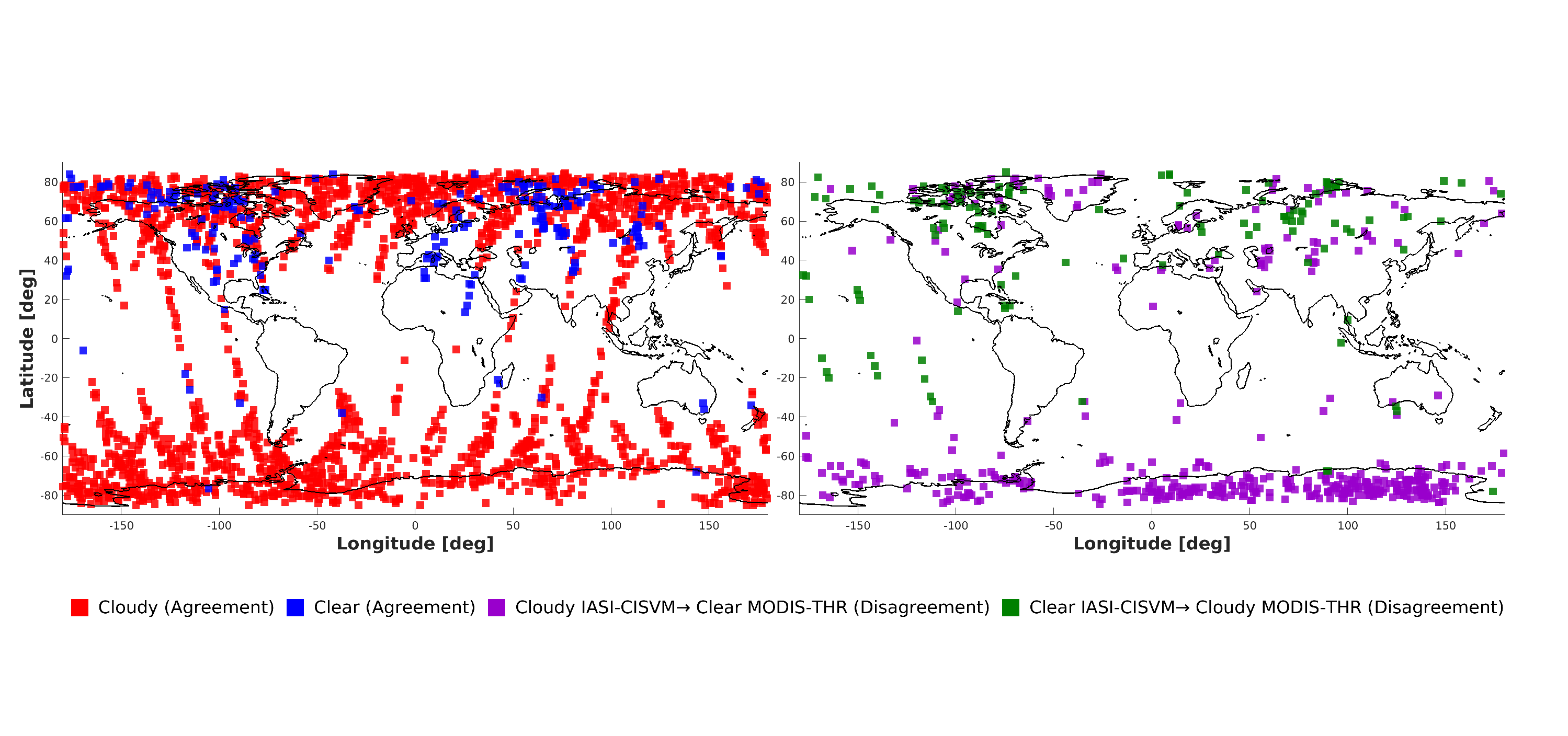}
    \caption{Differences in cloud detection agreement between the three cloud masks, at locations and times where all are available, for Summer. The figures in the panels are: IASI-THR vs MODIS-THR (top), IASI-THR vs IASI-CISVM (center), and IASI-CISVM vs MODIS-THR (bottom).}
    \label{fig:results_modis_svm_jul}
\end{figure}

Figure~\ref{fig:results_modis_svm_jan} and Figure~\ref{fig:results_modis_svm_jul} map the spatial patterns of agreement and disagreement for the three pairwise comparisons evaluated in Winter and Summer. As detailed in the figure captions, each panel compares two specific datasets. Points of mutual agreement are shown in red (both datasets detect clouds) and blue (both datasets detect clear sky). Areas of discrepancy are highlighted in violet (the first dataset detects clouds while the second detects clear sky) and green (the first dataset detects clear sky while the second detects clouds).

\begin{figure}[htbp]
    \centering
    \includegraphics[width=\textwidth]{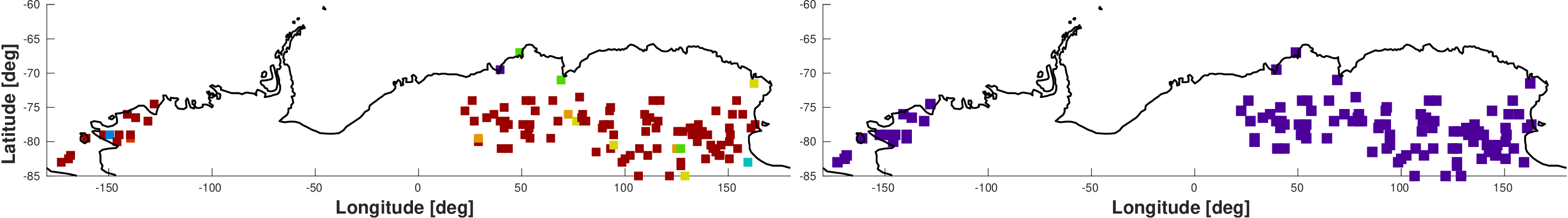}
    \vspace{0.2cm}
    \includegraphics[width=\textwidth]{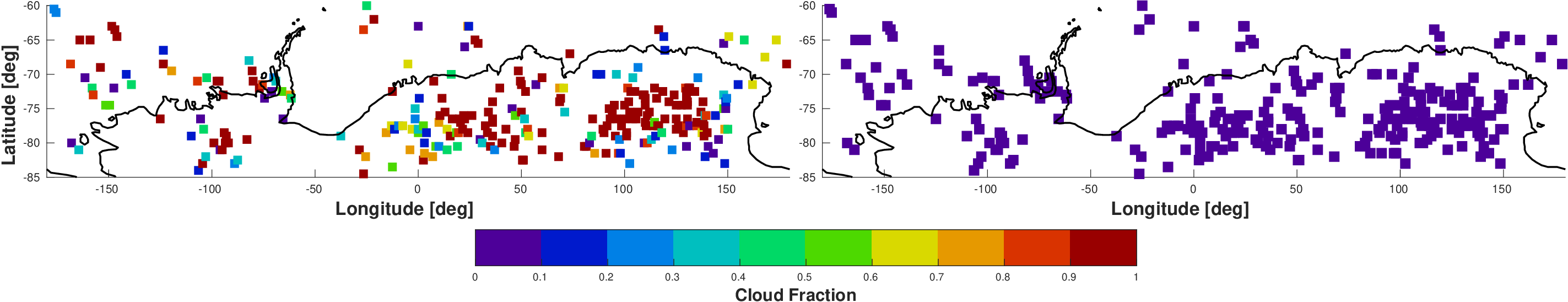}
    \caption{Cloud fraction over Antarctica corresponding to points classified as cloudy by IASI and clear by MODIS. The maps show the cloud fraction derived by IASI (left panels) and MODIS (right panels). Data for Winter are shown on top, and for Summer at the bottom.}
    \label{fig:cfraction_maps_antarctica_misclassified}
\end{figure}

\noindent It can be noted that the agreement between IASI and MODIS sensors is quite good in the tropical and temperate zones, while the agreement decreases in the polar zones. We report the original cloud fraction maps from IASI and MODIS for Antarctica in Figure~\ref{fig:cfraction_maps_antarctica_misclassified}. IASI detects a significantly higher cloud fraction, while MODIS reports clearer skies over the same locations and times. This discrepancy reflects fundamental differences in sensor sensitivity to polar radiative conditions. 

\noindent Cloud detection over Antarctica is particularly challenging because ice clouds and the snow-covered surface share similar radiative properties, making it difficult for satellite sensors to distinguish clear from cloudy scenes. MODIS often underestimates clouds in this region~\cite{Holz2008, Liu2010, volonnino2023}, especially during the polar night, owing to the limited availability of visible and near-infrared channels, very low surface temperatures, and frequent temperature inversions that reduce cloud surface thermal contrast; as a result, many cloudy scenes are classified as clear.

In contrast, the operational cloud product provided within the IASI L1C dataset, which is derived from collocated AVHRR measurements, tends to overdetect clouds over Antarctica~\cite{Whitburn_cloud_mask, Donat2025}. Under the thermal inversions common in this zone, imager-based cloud screening algorithms frequently struggle to separate highly emissive cold surfaces from clouds, which can cause clear-sky scenes to be misclassified as cloudy. This tendency confirms the broader difficulty of cloud detection in polar regions for passive sensors.

Beyond these surface induced false alarms, the detection capability from IASI radiances faces a different, yet equally significant, limitation: the underdetection of optically thin clouds. While optically thick clouds are generally well identified, detection performance naturally degrades as the radiance difference between clear and cloudy skies becomes minimal. FORUM simulations~\cite{sgheri2022} indicate that clouds with an optical depth below 0.03 at 900~\si{cm^{-1}} are generally undetectable, rising to 0.05 when only the mid-infrared band is considered.

Polar regions can also affect the detection of thicker clouds. Extremely low surface temperatures reduce the thermal contrast between the surface and cloud top, sometimes causing underdetection (as for MODIS) or overdetection (as for the AVHRR-derived reference).

Although the comparison with MODIS provides an important external consistency assessment, it does not remove the intrinsic limitation associated with the supervised training strategy adopted in this study. Since CISVM is trained using the operational IASI cloud product derived from collocated AVHRR observations, the classifier inevitably reflects, at least in part, the decision structure embedded in that operational reference. Consequently, systematic uncertainties affecting the reference cloud mask, for example over snow- and ice-covered surfaces or under weak cloud surface thermal contrast, naturally propagate into the trained model.

Consistent with the latitude-dependent error analysis presented in Section~\ref{sec:svm_to_IASI} and Table~\ref{tab:seasonal_variability}, the disagreement between IASI-CISVM and MODIS-THR is not uniformly distributed across the globe, but is concentrated in polar regions, where snow- and ice-covered surfaces reduce cloud surface radiative contrast for both sensors and amplify inter-product differences. The present results should therefore be interpreted as demonstrating the capability of hyperspectral infrared radiances to reproduce the large-scale behaviour of a widely used operational cloud-screening product and to remain broadly consistent with an independent passive cloud mask, rather than as an absolute validation of cloud detection accuracy.

Within this broader context, future validation activities should include comparisons with consensus cloud products and active-sensor benchmarks, in order to provide a more stringent assessment of reference-label uncertainty and of the generalization capability of the method. In parallel, ongoing work is directed toward more structured machine learning frameworks for infrared cloud classification, including formulations that can move beyond the present binary setting and explicitly represent intermediate cloud-coverage conditions within the IASI field of view.
\section{Conclusions\label{sec:conclusions}}
In this study, we proposed the Cloud Identification Support Vector Machine (CISVM), a supervised framework for binary clear/cloudy classification from IASI Level~1C hyperspectral infrared radiances, trained against the AVHRR-derived operational cloud cover product. The analysis, conducted over four seasons and across different radiometric representations, dimensionality-reduction strategies, and geographic subdivisions, shows that the best performance is obtained using radiances combined with PCA in the global (GLOBE) configuration, reaching 88.52\% agreement with the operational IASI cloud product. Clouds affect the entire infrared spectrum, so retaining more spectral information improves class separability with negligible inference cost.

Beyond the overall agreement, the results provide a physically interpretable picture of the classification behaviour. Performance degradation is concentrated over polar environments, where reduced cloud surface radiative contrast decreases the separability between clear and cloudy spectra. CISVM is entirely data-driven and therefore not a physically informed machine-learning method. However, soil type was included in the input features because it systematically influences the spectrum and therefore improves classification accuracy.

The comparison with collocated MODIS cloud products should be interpreted as an external consistency assessment across independent passive sensors, rather than as an absolute cloud-truth benchmark. CISVM reproduces the main large-scale behaviour of the operational IASI cloud product, with discrepancies concentrated in high-latitude regions, consistent with the known limitations of passive cloud detection over snow- and ice-covered surfaces. Since CISVM is trained on the AVHRR-derived IASI cloud product, the maximum achievable agreement is inherently constrained by the quality of that supervisory reference. This limitation is not specific to CISVM, but is a general characteristic of supervised learning approaches based on operational retrieval products.

Within this framework, the present study should be regarded both as a quantitative assessment of infrared cloud screening from IASI radiances and as a methodological baseline for ongoing developments. By establishing a validated SVM-based framework, including systematic evaluation across seasons, surface types, and spectral representations, together with statistical significance testing and computational cost analysis, this study provides the reference pipeline, data splits, and evaluation protocol required for a rigorous comparison with alternative learning-based approaches.

Current work is directed toward more flexible machine-learning frameworks. As a first step in this direction, a dedicated follow-up study is underway to evaluate a MultiLayer Perceptron (MLP) trained under the same preprocessing pipeline and evaluation protocol established here, in order to directly and fairly contextualize SVM performance against this alternative approach. This staged strategy ensures that any future comparison is performed against a methodologically mature and thoroughly characterized baseline, rather than against a preliminary or undervalidated reference.

In particular, future developments may investigate multi-class formulations able to distinguish clear, partially cloudy, and cloudy scenes, or more generally to represent intermediate cloud-coverage conditions within the IASI field of view. Within this broader context, future validation activities will include comparisons with cloud products derived from active sensors, in order to provide a more stringent assessment of reference label uncertainty and of the generalization capability of the method.

Overall, this study demonstrates that hyperspectral infrared radiances alone contain sufficient information to support physically interpretable and operationally relevant cloud screening at global scale, providing a robust methodological baseline for future infrared sounding missions lacking dedicated multispectral imagers.

\section*{Declarations}

\subsection*{Funding}
NRRP (National Recovery and Resilience Plan) supported all authors under the project EMM (Earth-Moon-Mars, Mission 4, Component 2, Investment 3.1, Project IR000038, CUPC53C22000870006).

\subsection*{CRediT authorship contribution statement}
\noindent{\bf Chiara Zugarini:} Methodology, Formal Analysis, Testing, Software, Writing – original draft, Writing – review \& editing
\noindent{\bf Cristina Sgattoni:} Formal Analysis, Testing, Software, Writing – original draft, Writing – review \& editing
\noindent{\bf Luca Sgheri:} Methodology, Formal Analysis, Writing – original draft, Writing – review \& editing, Funding Acquisition;

\subsection*{Conflict of interest}
The authors declare no conflict of interest.

 
\bibliographystyle{elsarticle-num}
\bibliography{references.bib}

\end{document}


\begin{frontmatter}

\title{Supplementary Material for\\
Machine Learning for Cloud Detection in IASI Measurements:
A Data-Driven SVM Approach with Physical Constraints}

\author[3,1]{Chiara Zugarini \orcidlink{0009-0005-8053-0646}\corref{cor1}}
\ead{chiara.zugarini@unifi.it}

\author[2]{Cristina Sgattoni \orcidlink{0000-0001-5734-0856}}
\ead{cristina.sgattoni@cnr.it}

\author[1]{Luca Sgheri \orcidlink{0000-0002-6014-9363}}
\ead{luca.sgheri@cnr.it}

\cortext[cor1]{Corresponding author}

\affiliation[1]{
    organization={CNR-IAC},
    addressline={Via Madonna del Piano, 10},
    city={Sesto Fiorentino},
    postcode={I-50019},
    state={FI},
    country={Italy}
}

\affiliation[2]{
    organization={CNR-IBE},
    addressline={Via Madonna del Piano, 10},
    city={Sesto Fiorentino},
    postcode={I-50019},
    state={FI},
    country={Italy}
}

\affiliation[3]{
    organization={Università degli Studi di Firenze, Dipartimento di Ingegneria dell'Informazione},
    addressline={Via Santa Marta, 3},
    city={Firenze},
    postcode={I-50139},
    state={FI},
    country={Italy}
}

\end{frontmatter}

\tableofcontents
\newpage


\section{Supplementary Material for Section 3: Creation of Training and Test Sets from IASI Measurements}
\label{sec:supp_trainingset_testset}

\subsection{IASI training set}\label{subsec:trainingset}

\begin{figure}[H]
\centering
\includegraphics[width=0.95\linewidth]{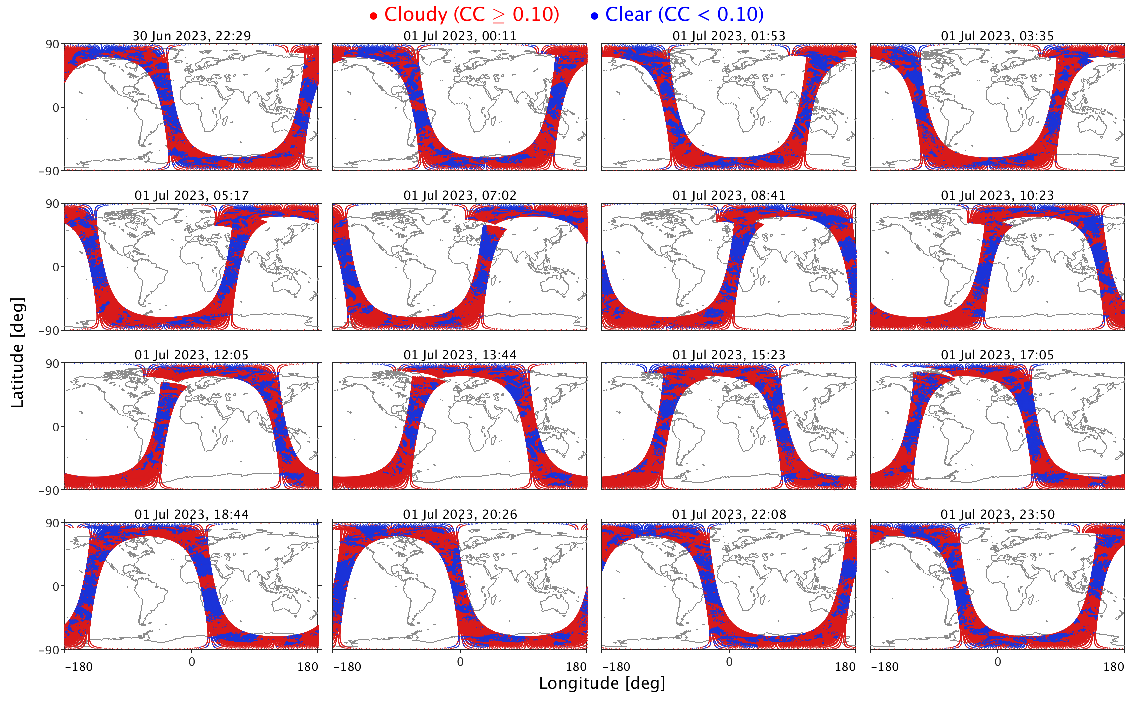}
\caption{Spatial distribution of IASI clear-sky (blue) and cloudy (red) observations for the Summer season, aggregated over the 16 training orbits listed in Table~\ref{tab:Orbits_training}.}
\label{sfig:IASI_cle_clo_JUL}
\end{figure}

\begin{figure}[H]
\centering
\includegraphics[width=0.95\linewidth]{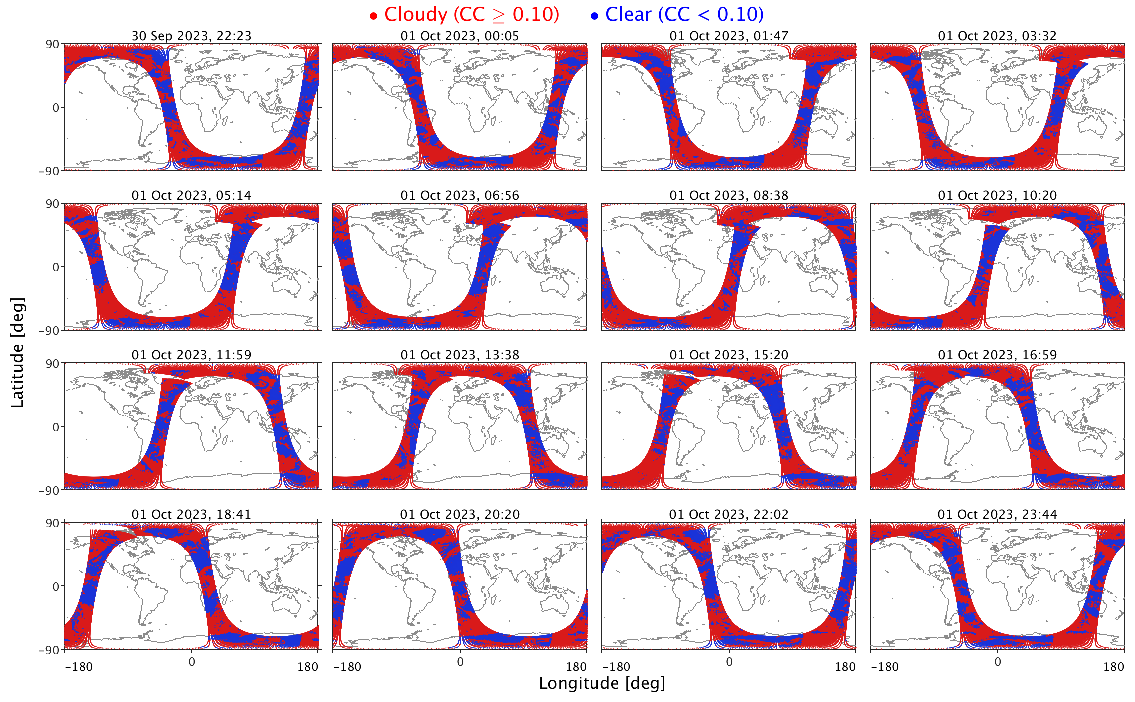}
\caption{Spatial distribution of IASI clear-sky (blue) and cloudy (red) observations for the Autumn season, aggregated over the 16 training orbits listed in Table~\ref{tab:Orbits_training}.}
\label{sfig:IASI_cle_clo_OCT}
\end{figure}

\begin{figure}[H]
\centering
\includegraphics[width=0.95\linewidth]{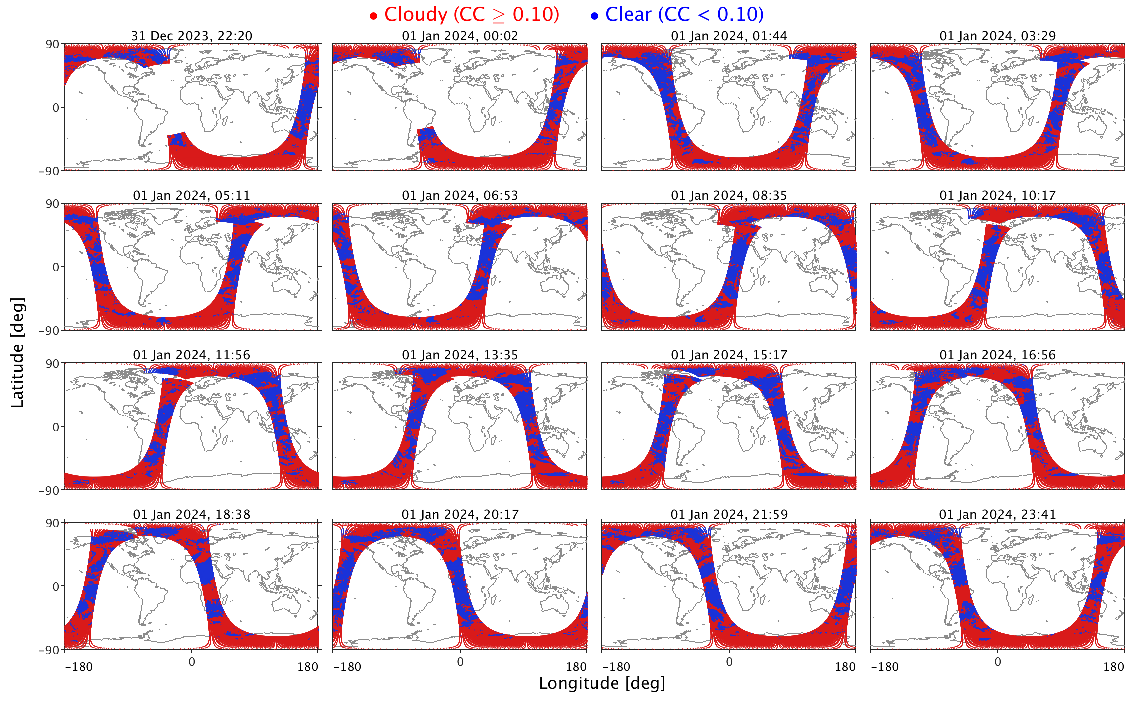}
\caption{Spatial distribution of IASI clear-sky (blue) and cloudy (red) observations for the Winter season, aggregated over the 16 training orbits listed in Table~\ref{tab:Orbits_training}.}
\label{sfig:IASI_cle_clo_JAN}
\end{figure}

Table~\ref{tab:Orbits_training} reports the complete list of the 64 IASI orbits used to construct the seasonal training datasets, 16 orbits per season, described in Section~3.1 of the main text. The spatial distribution of clear and cloudy IASI observations for the training set for Summer, Autumn, and Winter seasons, each aggregated over the 16 corresponding orbits, is shown in Figures~\ref{sfig:IASI_cle_clo_JUL},~\ref{sfig:IASI_cle_clo_OCT},~\ref{sfig:IASI_cle_clo_JAN}.

\begin{table}[H]
    \centering
    \resizebox{\textwidth}{!}{%
    \begin{tabular}{|c|c|c|c|c|c|}
        \hline
        \multicolumn{3}{|c|}{\textbf{Spring}} & \multicolumn{3}{c|}{\textbf{Summer}} \\
        \hline
        \textbf{Start Date} & \textbf{Start Time} & \textbf{Orbit ID} 
        & \textbf{Start Date} & \textbf{Start Time} & \textbf{Orbit ID} \\
        \hline
        2023-03-31 & 22:11:53 & 22815 & 2023-06-30 & 22:29:57 & 24108 \\
        \hline
        \textbf{End Date} & \textbf{End Time} & \textbf{Orbit ID} 
        & \textbf{End Date} & \textbf{End Time} & \textbf{Orbit ID} \\
        \hline
        2023-04-01 & 23:32:57 & 22830 & 2023-07-01 & 23:50:53 & 24123 \\
        \hline
        \multicolumn{3}{|c|}{Total Number of Orbits: 16} & \multicolumn{3}{c|}{Total Number of Orbits: 16} \\
        \hline
        \multicolumn{3}{|c|}{\textbf{Autumn}} & \multicolumn{3}{c|}{\textbf{Winter}} \\
        \hline
        \textbf{Start Date} & \textbf{Start Time} & \textbf{Orbit ID} 
        & \textbf{Start Date} & \textbf{Start Time} & \textbf{Orbit ID} \\
        \hline
        2023-09-30 & 22:23:54 & 25415 & 2023-12-31 & 22:20:54 & 26722 \\
        \hline
        \textbf{End Date} & \textbf{End Time} & \textbf{Orbit ID} 
        & \textbf{End Date} & \textbf{End Time} & \textbf{Orbit ID} \\
        \hline
        2023-10-01 & 23:44:58 & 25430 & 2024-01-01 & 23:41:58 & 26737 \\
        \hline
        \multicolumn{3}{|c|}{Total Number of Orbits: 16} & \multicolumn{3}{c|}{Total Number of Orbits: 16} \\
        \hline
    \end{tabular}%
   }
    \caption{Summary of starting and ending IASI orbits used for the training set in Spring, Summer, Autumn, and Winter. The orbits are daily and consecutive, spanning an acquisition period of approximately 24 hours for each season.}
    \label{tab:Orbits_training}
\end{table}

\subsection{IASI test set}\label{subsec:testset}

\begin{figure}[H]
\centering
\includegraphics[width=0.95\linewidth]{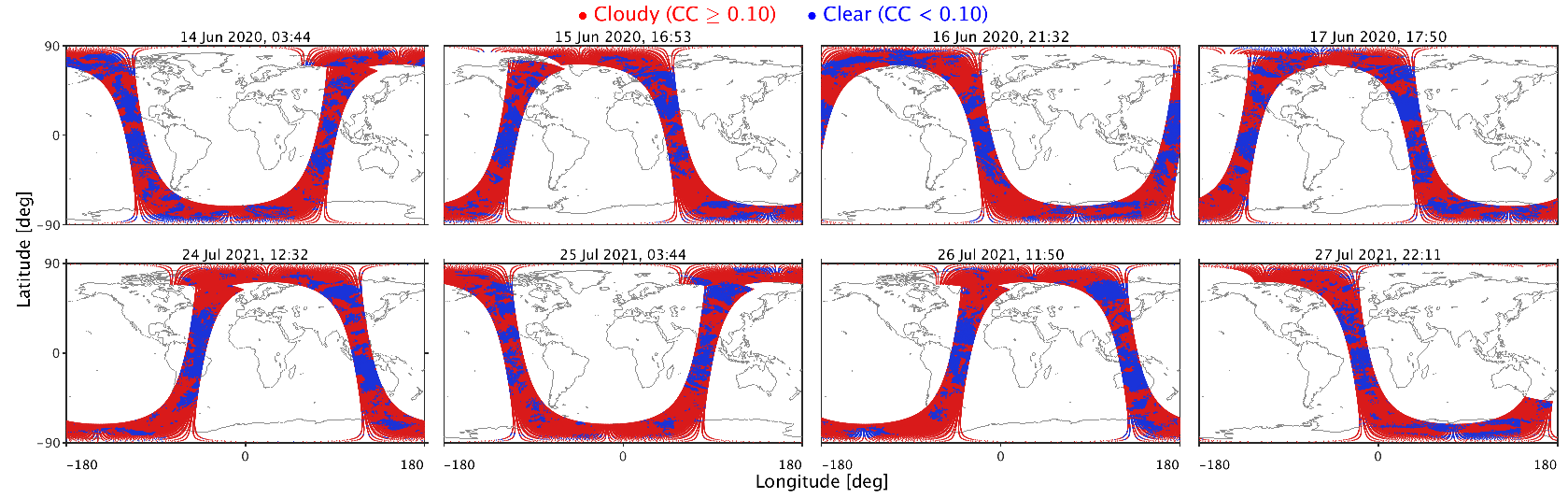}
\caption{Spatial distribution of IASI clear-sky (blue) and cloudy (red) observations for the Summer season, aggregated over the 8 test orbits listed in Table~\ref{tab:Orbits_test}.}
\label{sfig:IASI_cle_clo_testset_JUL}
\end{figure}

\begin{figure}[H]
\centering
\includegraphics[width=0.95\linewidth]{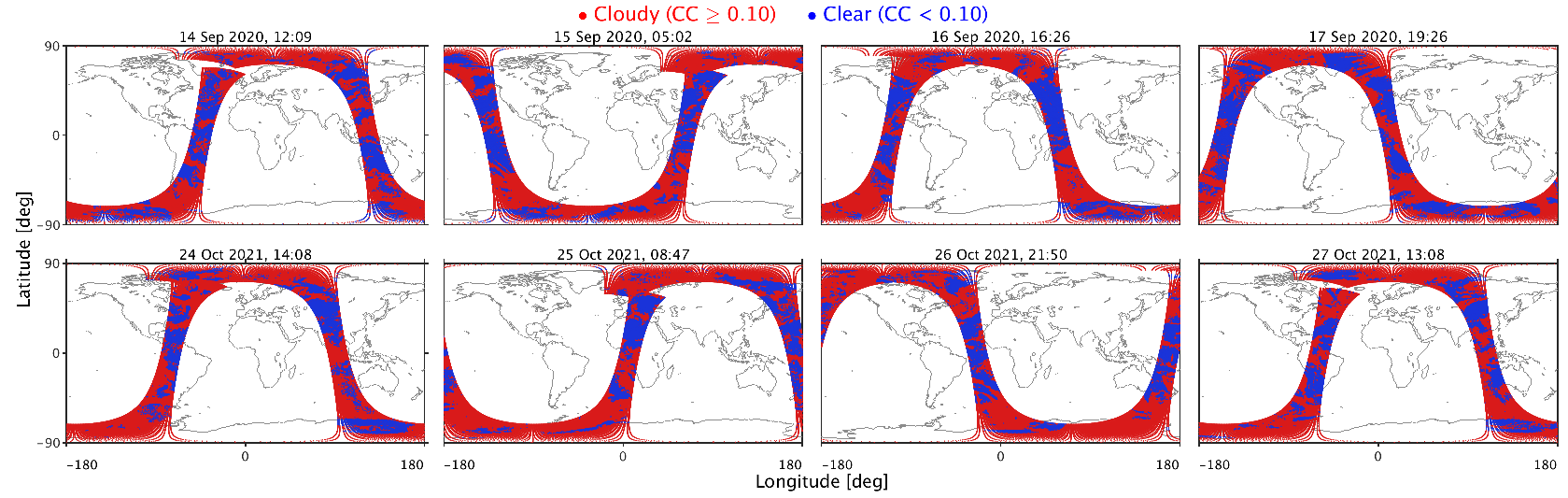}
\caption{Spatial distribution of IASI clear-sky (blue) and cloudy (red) observations for the Autumn season, aggregated over the 8 test orbits listed in Table~\ref{tab:Orbits_test}.}
\label{sfig:IASI_cle_clo_testset_OCT}
\end{figure}

\begin{figure}[H]
\centering
\includegraphics[width=0.95\linewidth]{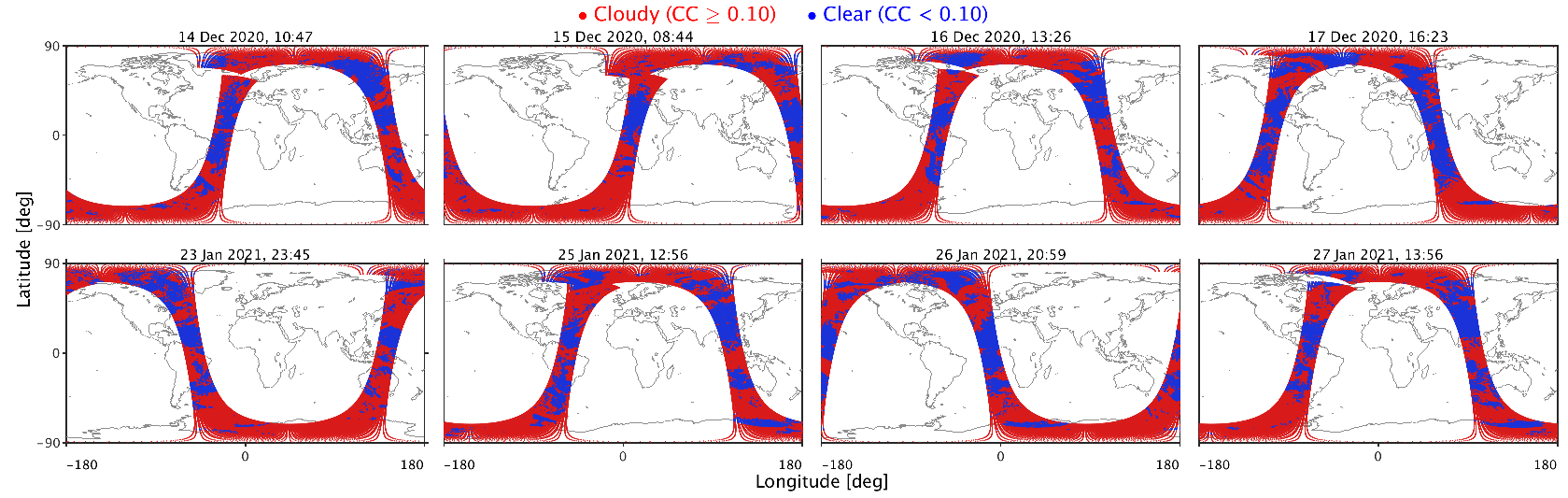}
\caption{Spatial distribution of IASI clear-sky (blue) and cloudy (red) observations for the Winter season, aggregated over the 8 test orbits listed in Table~\ref{tab:Orbits_test}.}
\label{sfig:IASI_cle_clo_testset_JAN}
\end{figure}

Table~\ref{tab:Orbits_test} reports the complete list of the 32 IASI orbits used to construct the independent seasonal test sets, 8 orbits per season, described in Section~3.2 of the main text. The spatial distribution of clear and cloudy IASI observations for the test set in Summer, Autumn, and Winter seasons, each aggregated over the 8 corresponding orbits, is shown in Figures~\ref{sfig:IASI_cle_clo_testset_JUL},~\ref{sfig:IASI_cle_clo_testset_OCT},~\ref{sfig:IASI_cle_clo_testset_JAN}.

\begin{table}[H]
    \centering
    \resizebox{\textwidth}{!}{%
    \begin{tabular}{|c|c|c|c|c|c|}
    \hline
    \multicolumn{3}{|c|}{\textbf{Spring}} & \multicolumn{3}{c|}{\textbf{Summer}} \\
    \hline
    \textbf{Date} & \textbf{Starting hour} & \textbf{Orbit ID} & \textbf{Date} & \textbf{Starting hour} & \textbf{Orbit ID} \\
    \hline
    2020-03-14 & 22:20:58 & 7017 & 2019-06-14 & 03:44:55 & 8313 \\
    \hline
    2020-03-15 & 13:35:54 & 7026 & 2019-06-15 & 16:53:59 & 8335 \\
    \hline
    2020-03-16 & 15:56:58 & 7044 & 2019-06-16 & 21:32:55 & 8352 \\
    \hline
    2020-03-17 & 17:53:54 & 7057 & 2021-06-17 & 17:50:55 & 8364 \\
    \hline
    2021-04-24 & 13:56:57 & 12780 & 2021-07-24 & 12:32:53 & 14072 \\
    \hline
    2021-04-25 & 13:35:52 & 12794 & 2021-07-25 & 03:44:53 & 14081 \\
    \hline
    2021-04-26 & 14:56:56 & 12809 & 2021-07-26 & 11:50:53 & 14100 \\
    \hline
    2021-04-27 & 16:14:56 & 12824 & 2021-07-27 & 22:11:17 & 14120 \\
    \hline
    \multicolumn{3}{|c|}{Total Number of Orbits: 8} & \multicolumn{3}{c|}{Total Number of Orbits: 8} \\
    \hline
    \multicolumn{3}{|c|}{\textbf{Autumn}} & \multicolumn{3}{c|}{\textbf{Winter}} \\
    \hline
    \textbf{Date} & \textbf{Starting hour} & \textbf{Orbit ID} & \textbf{Date} & \textbf{Starting hour} & \textbf{Orbit ID} \\
    \hline
    2020-09-14 & 12:09:00 & 9625 & 2020-12-14 & 10:47:52 & 10917 \\
    \hline
    2020-09-15 & 05:02:59 & 9635 & 2020-12-15 & 08:44:56 & 10930 \\
    \hline
    2020-09-16 & 16:26:59 & 9656 & 2020-12-16 & 13:26:56 & 10947 \\
    \hline
    2020-09-17 & 19:26:59 & 9672 & 2020-12-17 & 16:23:52 & 10963 \\
    \hline
    2021-10-24 & 14:08:57 & 15380 & 2021-01-24 & 23:45:00 & 11493 \\
    \hline
    2021-10-25 & 08:47:53 & 15391 & 2021-01-25 & 12:56:59 & 11515 \\
    \hline
    2021-10-26 & 21:50:57 & 15413 & 2021-01-26 & 20:59:55 & 11534 \\
    \hline
    2021-10-27 & 13:08:57 & 15422 & 2021-01-27 & 13:56:59 & 11544 \\
    \hline
    \multicolumn{3}{|c|}{Total Number of Orbits: 8} & \multicolumn{3}{c|}{Total Number of Orbits: 8} \\
    \hline
    \end{tabular}%
    }
    \caption{Selected IASI orbits for the test set in Spring, Summer, Autumn, and Winter.}
    \label{tab:Orbits_test}
\end{table}

\section{Supplementary Material for Section 4: Application of the algorithm to IASI measurements}\label{sec:svm_to_IASI}

\subsection{Support Vector Machine Formulation and Kernel Choice}\label{subsec:supp_svm_theory}

Formally, given a training dataset of $N$ labeled samples $\{(\mathbf{x}_i, y_i)\}_{i=1}^N$ where $\mathbf{x}_i \in \mathbb{R}^d$ and $y_i \in \{-1, 1\}$, a soft-margin SVM solves the convex optimization problem
\begin{equation}\label{eq:1}
\min_{\mathbf{w}, b, \boldsymbol{\xi}} \ \frac{1}{2} |\mathbf{w}|^2 + C \sum_{i=1}^N \xi_i,
\end{equation}
subject to
\begin{equation}\label{eq:2}
y_i \bigl(\mathbf{w}^\top \phi(\mathbf{x}_i) + b\bigr) \geq 1 - \xi_i, \quad \xi_i \geq 0, \quad \forall i,
\end{equation}
where $\mathbf{w} \in \mathcal{H}$ is the normal vector of the separating hyperplane, $b \in \mathbb{R}$ is the bias term, and $\phi:\mathbb{R}^d \to \mathcal{H}$ is a (possibly nonlinear) mapping into a higher-dimensional feature space $\mathcal{H}$. Here $d$ denotes the dimensionality of the input feature space (e.g., the number of spectral channels or principal components), while $\mathcal{H}$ may have a much higher, possibly infinite, dimensionality depending on the kernel choice. For non-linearly separable data, slack variables $\xi_i \ge 0$ are introduced to allow margin violations, and the regularization parameter $C > 0$ controls the trade-off between maximizing the margin and minimizing classification errors.

Introducing Lagrange multipliers leads to the dual formulation
\begin{equation}\label{eq:4}
\max_{\boldsymbol{\alpha}} \ \sum_{i=1}^N \alpha_i - \frac{1}{2} \sum_{i=1}^N \sum_{j=1}^N \alpha_i \alpha_j y_i y_j K(\mathbf{x}_i, \mathbf{x}_j),
\end{equation}
subject to
\begin{equation}\label{eq:5}
\sum_{i=1}^N \alpha_i y_i = 0, \quad 0 \leq \alpha_i \leq C, \quad \forall i,
\end{equation}
where $K(\mathbf{x}_i,\mathbf{x}_j)=\langle \phi(\mathbf{x}_i),\phi(\mathbf{x}_j)\rangle$ is a positive-definite kernel function that implicitly computes inner products in $\mathcal{H}$. Once the optimal multipliers $\alpha_i$ are obtained, the decision function is expressed as
\begin{equation}\label{eq:3}
f(\mathbf{x}) = \operatorname{sign} \left( \sum_{i=1}^N \alpha_i y_i K(\mathbf{x}_i, \mathbf{x}) + b \right),
\end{equation}
eliminating the need for an explicit feature mapping $\phi$, since inner products in $\mathcal{H}$ are computed through $K$ (the kernel trick).

In this work, we consider three standard kernels:
\begin{itemize}
\item[-] \textbf{Linear kernel:} $K(\mathbf{x}_i, \mathbf{x}_j) = \mathbf{x}_i^\top \mathbf{x}_j$,
\item[-] \textbf{Polynomial kernel:} $K(\mathbf{x}_i, \mathbf{x}_j) = (\mathbf{x}_i^\top \mathbf{x}_j + c)^d$,
\item[-] \textbf{Gaussian (RBF) kernel:} $K(\mathbf{x}_i, \mathbf{x}_j) = \exp(-\gamma |\mathbf{x}_i - \mathbf{x}_j|^2)$.
\end{itemize}
The polynomial kernel is particularly effective when the input features are correlated: by raising the inner product \( \mathbf{x}^\top \mathbf{x}' \) to the power \( d \), it leverages feature correlations in a higher-dimensional space and enhances class separability. In contrast, the Gaussian (RBF) kernel is more suitable for data with weaker global correlations, acting as a localized similarity measure that emphasizes proximity in the input space and mapping samples into an infinite-dimensional feature space where highly non-linear decision boundaries can be represented.

To determine the most appropriate kernel for our application, we rely on the analysis in~\cite{zugarini2023}, which investigated kernel choice for cloud detection using simulated FORUM hyperspectral data~\cite{rfms19, palchetti2020, sgheri2022, ridolfi20} and showed that the polynomial kernel outperforms alternative kernels for radiances characterized by strong inter-channel correlations. This is consistent with the intrinsic spectral structure of instruments such as IASI, where adjacent wavenumbers are highly correlated. Consequently, in our experiments over continuous spectral ranges we adopt the polynomial kernel, whereas for classifications based on isolated and weakly correlated channels the Gaussian kernel is preferred. SVM models are implemented using the \texttt{fitcsvm} function~\cite{matlab1, matlab2} in the MATLAB environment, which provides built-in routines for training and cross-validating binary SVM classifiers.

\subsection{Brightness temperature}\label{subsec:brightness_temperature}
In this work, spectral radiances can be expressed as brightness temperatures (BT), obtained by inverting Planck's law. The brightness temperature at wavenumber $\nu$ is defined as

\begin{equation}
\label{eq:BT}
T_b(\nu)=\frac{hc\nu}{k}
\left[
\ln\left(
\frac{2hc^2\nu^3}{R(\nu)}+1
\right)
\right]^{-1},
\end{equation}

where

\begin{itemize}
    \item $T_b(\nu)$ is the brightness temperature at wavenumber $\nu$ [K];
    \item $\nu$ is the wavenumber [cm$^{-1}$];
    \item $R(\nu)$ is the spectral radiance at wavenumber $\nu$ [W\,m$^{-2}$\,sr$^{-1}$\,(cm$^{-1}$)$^{-1}$];
    \item $h$ is Planck's constant ($6.626\times10^{-34}$ J\,s);
    \item $c$ is the speed of light in vacuum ($2.998\times10^{8}$ m\,s$^{-1}$);
    \item $k$ is Boltzmann's constant ($1.381\times10^{-23}$ J\,K$^{-1}$).
\end{itemize}

The brightness temperature represents the temperature a blackbody would need in order to emit the same spectral radiance observed at a given wavenumber. In simple terms, it provides a way to normalize radiance with respect to the Planck function, facilitating physical interpretation.

\subsection{Preliminary Spectral Band Selection Test}\label{subsec:preliminary_test}
A preliminary identification test based on radiance data was carried out to identify the most effective spectral subsetting strategy for establishing a baseline against the PCA approach. This preliminary test incorporated the geographic and soil-type stratifications described in Subsection~3.3, and was restricted to the two extreme seasons (Summer and Winter) to capture maximum variability.

Tests A to G evaluated consecutive wavenumber bands within the main IASI spectral regions~\cite{esa_metop_spectral_range}, partitioned according to their primary atmospheric and surface applications. Test H employed the specific channel selection proposed by Whitburn et al.~\cite{Whitburn_cloud_mask}, comprising 45 discrete IASI channels highly sensitive to multi-layer cloud properties.

The primary objective of this assessment was to evaluate the capability of the SVM classifier to exploit raw radiance information across physically meaningful spectral domains, prior to any dimensionality reduction. The kernel functions were selected based on the correlation structure of the respective input spaces: a polynomial kernel was applied to Tests A to G to effectively handle the strong mutual correlation inherent in contiguous spectral bands, while a Gaussian (RBF) kernel was adopted for Test H, where the individually selected, dispersed channels exhibit minimal mutual correlation, making an RBF decision boundary more suitable for the high-dimensional feature space.

\begin{table}[H]
    \centering
    \resizebox{\textwidth}{!}{%
    \begin{tabular}{
        |>{\centering\arraybackslash}p{1cm}
        |>{\centering\arraybackslash}p{4.3cm}
        |>{\centering\arraybackslash}p{4.3cm}
        |>{\centering\arraybackslash}p{3.2cm}
        |>{\centering\arraybackslash}p{3.2cm}|
    }
        \hline
        \textbf{Test} & \textbf{Spectral Range or Selected Channels [cm$^{-1}$]} & \textbf{Primary Application} &
        \begin{tabular}[c]{@{}c@{}}\\ \textbf{Total Rate}\\ \textbf{Subdivision (a)}\\ \small N. Tests: 204612\end{tabular} &
        \begin{tabular}[c]{@{}c@{}}\\ \textbf{Total Rate}\\ \textbf{Subdivision (b)}\\ \small N. Tests: 204612\end{tabular} \\
        \hline\hline
        A & [645; 2760] & IASI range~\cite{esa_metop_spectral_range} 
        & \begin{tabular}{c}107844\\52.70\%\end{tabular} 
        & \begin{tabular}{c}105372\\51.50\%\end{tabular} \\
        \hline
        B & [800; 900] & Atmospheric Window 
        & \begin{tabular}{c}113884\\55.66\%\end{tabular} 
        & \begin{tabular}{c}125376\\61.27\%\end{tabular} \\
        \hline
        C & [770; 980] & Surface and cloud properties~\cite{esa_metop_spectral_range} 
        & \begin{tabular}{c}121684\\59.47\%\end{tabular} 
        & \begin{tabular}{c}136384\\66.65\%\end{tabular} \\
        \hline
        D & [1080; 1150] & Surface and cloud properties~\cite{esa_metop_spectral_range} 
        & \begin{tabular}{c}136808\\66.86\%\end{tabular} 
        & \begin{tabular}{c}141520\\69.17\%\end{tabular} \\
        \hline
        E & [2420; 2700] & Surface and cloud properties~\cite{esa_metop_spectral_range} 
        & \begin{tabular}{c} 97196\\47.50\%\end{tabular} 
        & \begin{tabular}{c} 119488\\58.40\%\end{tabular} \\
        \hline
        F & [770; 980] & Surface and cloud properties~\cite{esa_metop_spectral_range} 
        & \begin{tabular}{c}123612\\60.41\%\end{tabular} 
        & \begin{tabular}{c}131248\\64.14\%\end{tabular} \\
          & \vspace{-0.8cm}[1080; 1150] & & & \\
        \hline
        
        G & [770; 980] & Surface and cloud  
        & \begin{tabular}{c} 132872\\64.94\%\end{tabular} 
        & \begin{tabular}{c} 133080\\65.04\%\end{tabular} \\
        
          & \vspace{-0.5cm}[1080; 1150] & properties~\cite{esa_metop_spectral_range} & & \\
          
          & \vspace{-0.4cm}[2420; 2700] & & & \\
          
        \hline
        H & [826, 827.50, 861.50, 865.50, 866.25, 869.25, 871.25, 874.75, 877, 878.50, 880, 883.75, 885.75, 887.25, 891.25, 894.25, 897.75, 899.75, 901.50, 902.50, 905, 994.50, 996.25, 999.50, 1001.50, 1004.75, 1006.50, 1009.75, 1011.50, 1014.50, 2145.50, 2147.25, 2150, 2152.50, 2153.50, 2157.25, 2158.25, 2161.75, 2164.75, 2166.75, 2169.25, 2172.75, 2174.25, 2176.25, 2177.75] 
        & IASI range with 45 selected channels~\cite{Whitburn_cloud_mask} 
        & \begin{tabular}{c}157016\\76.74\%\end{tabular} 
        & \begin{tabular}{c}150732\\73.67\%\end{tabular} \\
        \hline
    \end{tabular}%
    }
    \caption{For each test type, the second and third columns report the selected spectral range or specific channels and their corresponding primary application. The last two columns show the total rate for subdivisions (a) and (b) as defined in Table~5, including the total number of correctly classified points and the corresponding success rate (\%) computed over the combined Winter and Summer test sets.}
    \label{tab:Spectral_range}
\end{table}
As reported in Table~\ref{tab:Spectral_range}, the highest overall accuracy is achieved by Test H ($76.74\%$ for Subdivision~(a)), which demonstrates a clear advantage over continuous-band selection methods (Tests A–G, ranging from $47.50\%$ to $69.17\%$). Based on this evidence, the Whitburn et al. selection scheme was adopted as the optimal fixed-channel baseline for the main comparative analysis.

\subsection{Threshold Sensitivity Analysis}\label{subsec:cc_sensitivity}

\begin{figure}[htbp]
    \centering
    \begin{subfigure}[b]{0.48\textwidth}
        \centering
        \includegraphics[width=\textwidth]{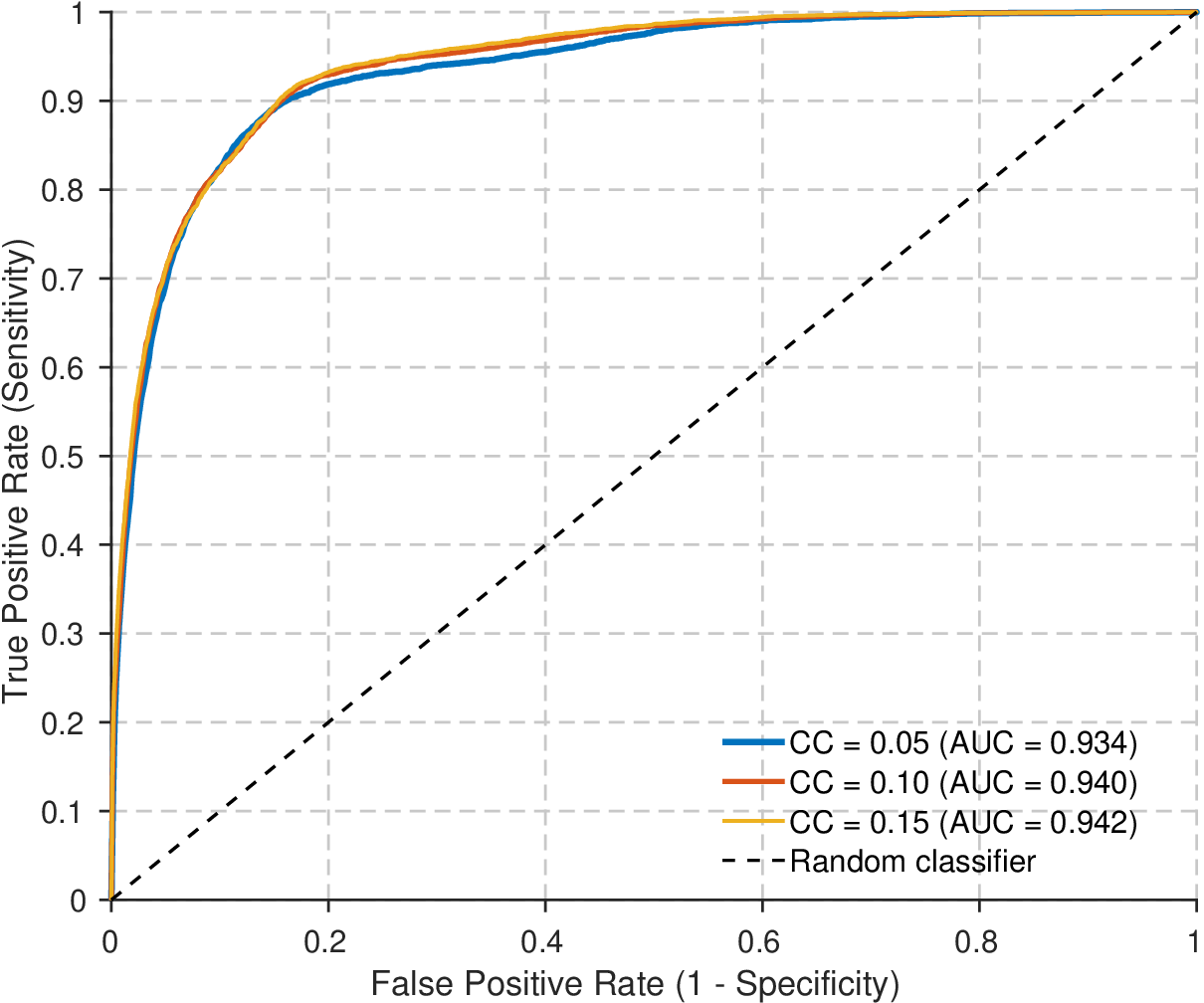}
        \caption{ROC curves}
        \label{fig:roc_sensitivity}
    \end{subfigure}
    \hfill
    \begin{subfigure}[b]{0.48\textwidth}
        \centering
        \includegraphics[width=\textwidth]{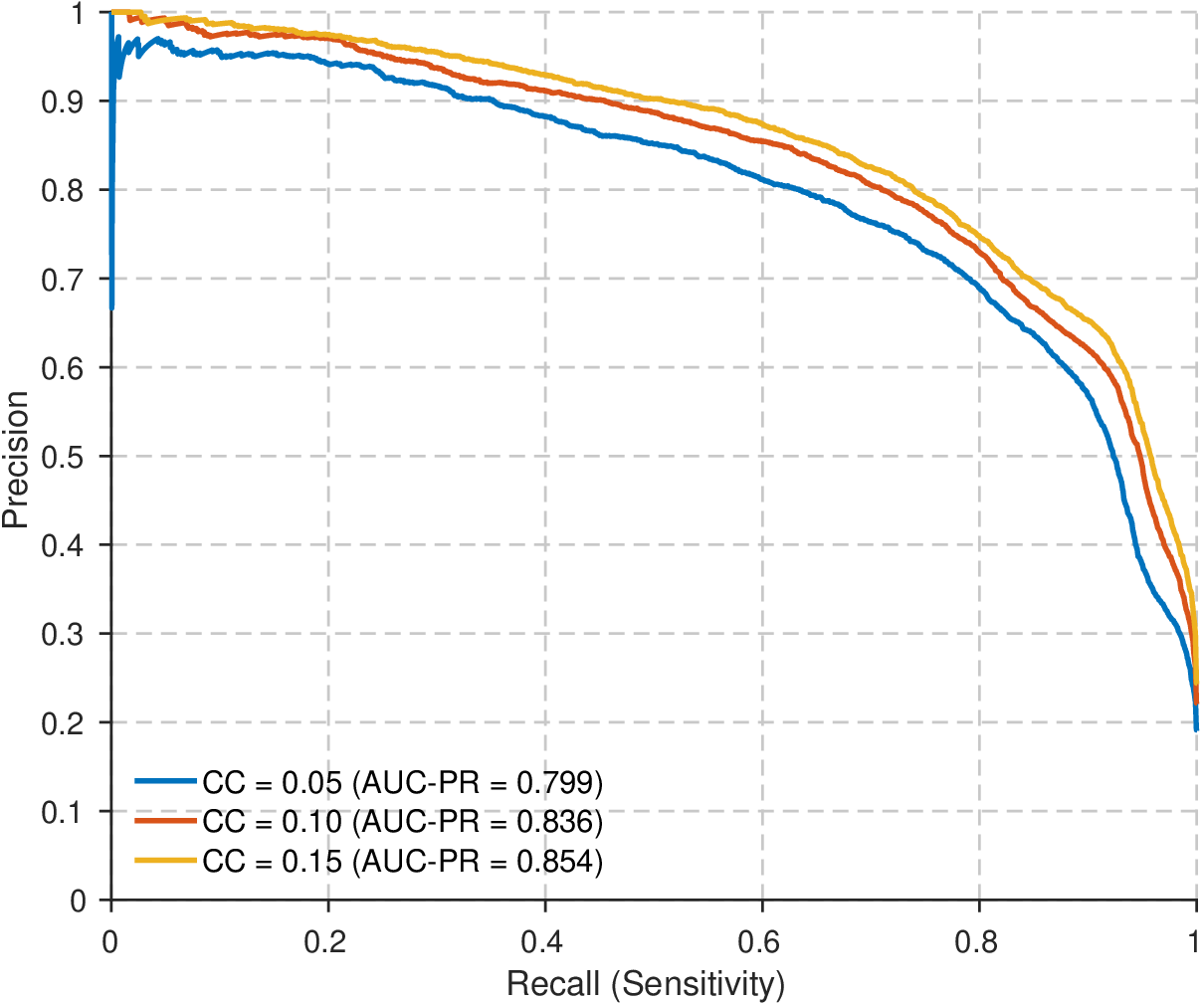}
        \caption{Precision-Recall curves}
        \label{fig:pr_sensitivity}
    \end{subfigure}
    \caption{ROC (a) and Precision-Recall (b) curves for the selected RAD-GLOBE-PCA configuration (Summer season), evaluated using three alternative CC labeling thresholds (0.05, 0.10, 0.15).}
    \label{fig:cc_sensitivity}
\end{figure}

To confirm the robustness of the selected RAD-GLOBE-PCA configuration, we evaluated its sensitivity to the clear and cloudy labeling threshold. Figure~\ref{fig:cc_sensitivity} shows the Receiver Operating Characteristic (ROC) and Precision-Recall (PR) curves obtained by varying the CC threshold (0.05, 0.10, and 0.15) for the Summer test set, pooling all surface types.

The classification performance proves to be largely independent of the adopted threshold. The ROC analysis yields nearly identical curves (AUC ranging from 0.934 to 0.942, variation $<1\%$), confirming a stable overall discrimination capability across labeling definitions. As shown by the PR curves, varying the threshold introduces a more visible modulation in the AUC--PR (from 0.799 to 0.854), which reflects the inherently stricter task of detecting weakly contaminated scenes at the lower threshold CC$=0.05$, where the positive class becomes more constrained and harder to separate. Taken together, these results demonstrate that the classifier's discrimination capability is not an artifact of the labeling definition, and support the use of the conservative CC$=0.10$ criterion, which offers a balanced operating point between the two extremes, for the subsequent surface-type analysis.

\subsection{PCA Configuration}\label{subsec:S_pca_configuration}

For the PCA-based configurations, dimensionality reduction was independently performed for each test using the corresponding training dataset generated according to the geographical subdivision and the training partitioning reported in Table~2 of the main text. Prior to the PCA projection, the spectral data were rigorously standardized to zero mean and unit variance. To prevent data leakage, the standardization of the test set was systematically performed by centering and scaling the data using the mean and standard deviation vectors computed exclusively from the corresponding training set. The number of retained principal components was automatically determined by imposing a cumulative explained variance threshold of 99.5\%, ensuring that the reduced feature space preserved most of the spectral information content associated with each spectrum.

\begin{table}[htbp]
\centering
\resizebox{\textwidth}{!}{%
\begin{tabular}{|c|c||c|c|c|c|}
\hline
\makecell{\textbf{Radiometric}\\\textbf{Input}} &
\makecell{\textbf{Geography}\\\textbf{Division}} &
\textbf{SPRING} &
\textbf{SUMMER} &
\textbf{AUTUMN} &
\textbf{WINTER} \\
\hline
\hline
BT & GLOBE &
\makecell{348 \\ {[99.50, 524.50]} \\ 4.11\%} &
\makecell{181.50 \\ {[33.00, 348.00]} \\ 2.15\%} &
\makecell{220 \\ {[63.50, 560.50]} \\ 2.60\%} &
\makecell{274 \\ {[91.50, 655.50]} \\ 3.24\%} \\
\hline
BT & ZONES &
\makecell{194 \\ {[78.75, 567.25]} \\ 2.29\%} &
\makecell{171 \\ {[69.75, 308.25]} \\ 2.02\%} &
\makecell{239 \\ {[49.00, 527.00]} \\ 2.82\%} &
\makecell{121 \\ {[27.00, 445.00]} \\ 1.43\%} \\
\hline
RAD & GLOBE &
\makecell{263 \\ {[123.50, 309.50]} \\ 3.11\%} &
\makecell{119 \\ {[24.00, 165.00]} \\ 1.41\%} &
\makecell{231 \\ {[104.50, 253.00]} \\ 2.73\%} &
\makecell{134 \\ {[53.00, 193.00]} \\ 1.58\%} \\
\hline
RAD & ZONES &
\makecell{202 \\ {[78.75, 344.25]} \\ 2.39\%} &
\makecell{187 \\ {[119.50, 325.50]} \\ 2.21\%} &
\makecell{218 \\ {[130.00, 403.00]} \\ 2.58\%} &
\makecell{166 \\ {[70.00, 299.00]} \\ 1.96\%} \\
\hline
\end{tabular}%
}
\caption{
Seasonal statistics of the PCA dimensionality for the different experimental configurations.
Reported values correspond to the median retained principal components, the IQR, and the median percentage of selected components, respectively.
}
\label{tab:pca_statistics}
\end{table}

Table~\ref{tab:pca_statistics} reports, for each configuration and season, the median number of retained principal components, the interquartile range (IQR) of the retained PCA dimensionality, and the corresponding median percentage of selected components. The median represents the typical number of retained principal components required to preserve the target spectral information content under the considered atmospheric conditions. The IQR represents the interval containing the central 50\% of the retained PCA dimensionality values and provides a robust estimate of the variability of the spectral complexity under different atmospheric conditions. The percentage of retained components represents the fraction of the principal components kept with respect to the original spectral dimensionality, providing a quantitative estimate of the achieved spectral reduction.

Global configurations tend to require a larger number of retained principal components, particularly for BT representations, indicating that globally distributed datasets encompass a wider range of atmospheric and radiative regimes. Seasonal differences in the retained PCA dimensionality suggest that the effective spectral complexity varies with the atmospheric state and radiative conditions. Despite the large original spectral dimensionality, the resulting percentages of retained components remain relatively low for all configurations, confirming the strong redundancy of spectral infrared observations.

\subsection{SVM Configuration}\label{subsec:S_svm_configuration}

All final CISVM configurations employ a Gaussian radial basis function (RBF) kernel. The SVM regularization parameter Box Constraint, denoted as $C$, was fixed to 1 for all tests in order to ensure a uniform regularization level and a fair comparison among the different configurations.

The Gaussian kernel introduces a characteristic similarity scale within the spectral feature space, controlling the spatial extent of the nonlinear decision boundaries generated by the classifier. Conceptually, this scale dictates the radius of influence of individual training samples: small \texttt{KernelScale} values produce highly localized, intricate boundaries to separate closely intertwined classes, whereas large values result in smoother, broader decision boundaries. In the MATLAB implementation, this similarity scale is governed by the parameter \texttt{KernelScale}, which is related to the Gaussian kernel parameter $\gamma$ through

\begin{equation}
\mathrm{KernelScale} =
\frac{1}{\sqrt{2\gamma}}.
\end{equation}

\noindent For all experimental configurations, the parameter $\gamma$ was estimated from the statistical properties of the corresponding training dataset according to

\begin{equation}
\gamma =
\frac{1}{N_f \sigma^2}
\end{equation}

\noindent where $N_f$ denotes the number of spectral features and $\sigma^2$ represents the global variance computed over the training dataset. This formulation allows the kernel similarity scale to automatically adapt to the dimensionality and statistical variability of the adopted spectral representation.

To characterize the geometry of the spectral feature space associated with the different atmospheric conditions and spectral representations, the Gaussian kernel scale was statistically analyzed for all test configurations and seasons.

\begin{table}[htbp]
\centering
\small
\begin{tabular}{|c||c|c|}
\hline
\textbf{Spectral Domain} &
\begin{tabular}{c}\textbf{Kernel Scale} \\ \textbf{(Mean $\pm$ Std)}\end{tabular} &
\begin{tabular}{c}\textbf{IQR Width} \\ \textbf{(Mean $\pm$ Std)}\end{tabular} \\
\hline
\hline 
CHANNELS & $0.660 \pm 0.096$ & $0.361 \pm 0.133$ \\
\hline
PCA & $64.872 \pm 0.004$ & $0.072 \pm 0.026$ \\
\hline
\end{tabular}
\caption{
Global statistics of the Gaussian kernel scale for the CHANNELS and PCA configurations.
The IQR width is defined as the difference between the third and first quartiles.
}
\label{tab:kernel_scale_summary}
\end{table}

Table~\ref{tab:kernel_scale_summary} summarizes the global statistical behavior of the Gaussian kernel scale for the CHANNELS and PCA spectral representations. The CHANNELS configurations exhibit relatively small kernel-scale values together with comparatively large variability, indicating that the optimal kernel scale is more dependent on the specific dataset configuration. The substantially reduced variability observed for the PCA-based configurations suggests that PCA reduces the dimensionality and removes low-variance spectral directions, leading to a more compact and consistent feature space.

\subsection{Surface-Type Dependence of Classification Performance}\label{subsec:S_surface_type_analysis}

This subsection provides the formal definitions of the classification metrics used throughout the main text to evaluate cloud-detection performance, together with a detailed analysis of their dependence on surface type.

Let $\mathrm{TP}_{c}$, $\mathrm{FP}_{c}$ and $\mathrm{FN}_{c}$ denote, respectively, the numbers of True Positives, False Positives and False Negatives for class $c \in \{\text{cloud},\text{clear}\}$, evaluated on an independent test set of size $N$. Using these components, the complete set of evaluation metrics adopted in this study is defined and detailed in Table~\ref{tab:metrics}.

\begin{table}[H]
\centering
\renewcommand{\arraystretch}{1.45}
\begin{tabular}{
@{}
p{0.13\textwidth}
p{0.33\textwidth}
p{0.44\textwidth}
@{}
}
\toprule
\textbf{Metric} &
\textbf{Formula} &
\textbf{Meaning}
\\
\midrule

Accuracy &
$\displaystyle
\frac{\mathrm{TP}_{\text{cloud}}+\mathrm{TP}_{\text{clear}}}{N}
$
&
Overall classification performance; fraction of correctly classified pixels.
\\[0.4em]

Cloud Recall &
$\displaystyle
\frac{\mathrm{TP}_{\text{cloud}}}
{\mathrm{TP}_{\text{cloud}}+\mathrm{FN}_{\text{cloud}}}
$
&
Sensitivity to cloudy scenes;\newline
fraction of truly cloudy pixels correctly detected.
\\[0.4em]

Clear\newline Recall &
$\displaystyle
\frac{\mathrm{TP}_{\text{clear}}}
{\mathrm{TP}_{\text{clear}}+\mathrm{FN}_{\text{clear}}}
$
&
Sensitivity to clear-sky scenes; fraction of truly clear-sky pixels correctly identified.
\\[0.4em]

Precision &
$\displaystyle
\frac{\mathrm{TP}_{\text{cloud}}}
{\mathrm{TP}_{\text{cloud}}+\mathrm{FP}_{\text{cloud}}}
$
&
Reliability of cloud detections; fraction of predicted cloudy pixels that are truly cloudy.
\\[0.4em]

Balanced Accuracy &
$\displaystyle
\frac{\mathrm{Recall}_{\text{cloud}}+
\mathrm{Recall}_{\text{clear}}}{2}
$
&
Class-balanced performance measure obtained by averaging Cloud Recall and Clear Recall.
\\[0.4em]

F1-score &
$\displaystyle
\frac{
2\,\mathrm{Recall}_{\text{cloud}}\,
\mathrm{Precision}_{\text{cloud}}
}{
\mathrm{Recall}_{\text{cloud}}+
\mathrm{Precision}_{\text{cloud}}
}
$
&
Harmonic mean of Cloud Recall and Precision; balanced measure of cloud-detection skill accounting for both omission and commission errors.
\\
\bottomrule
\end{tabular}
\caption{Summary of the classification metrics used to evaluate cloud detection performance.}
\label{tab:metrics}
\end{table}

The complexity of the trained SVM can be quantified through the Support Vector Ratio (SVRatio), defined as
\begin{equation}
\mathrm{SVRatio} = \frac{N_{\mathrm{SV}}}{N_{\mathrm{train}}},
\label{eq:svratio}
\end{equation}
where $N_{\mathrm{SV}}$ denotes the number of support vectors and $N_{\mathrm{train}}$ the number of training samples. In an SVM, support vectors include both margin-defining and margin-violating samples. Consequently, the SVRatio provides an empirical proxy for classification complexity: relatively low values generally indicate well-separated classes, whereas higher values imply increased overlap between cloudy and clear-sky observations, requiring a more complex decision boundary.

As shown in Figure~7 of the main text, non-land surfaces exhibit the lowest SVRatio ($0.20$), which is likely consistent with the homogeneity of water emissivity and skin temperature at the global scale. All land classes display substantially larger values (ranging from $0.29$ for Fine soils up to $0.43$ for Tropical Organic soils), suggesting an increased overlap between cloudy and clear-sky observations driven by surface spectral and thermal heterogeneity. These elevated values indicate that soil-dependent variations in emissivity and surface properties broaden the within-class distribution of clear-sky radiances and reduce their separability from cloudy observations.

Cross-referencing with Figure~5 reveals two distinct regimes of high complexity. For Coarse and Tropical Organic soils, elevated SVRatio values coincide with reduced Clear Recall and Balanced Accuracy, indicating that the increased cluster overlap likely translates into a degradation of classification performance. Conversely, for Very Fine soils, a high SVRatio ($0.37$) coexists with the highest F1-score ($93.6\%$) and Precision ($95.5\%$): here, it appears that while the classifier requires a more complex decision boundary, the underlying cloudy and clear-sky populations remain sufficiently separable to sustain excellent performance.

The variability of SVRatio across soil classes substantially exceeds its seasonal variability, suggesting that surface radiative properties exert a stronger control on classifier complexity than seasonal effects, and providing further indirect evidence that land-surface heterogeneity acts as the primary source of uncertainty in infrared cloud detection.

\subsection{Computational Cost and Memory Footprint}\label{subsec:S_computational_cost_and_memory}

For each test configuration described in Section~4 of the main text, the computational cost was evaluated separately for the SVM training phase and the prediction phase. Since the individual geographical subdivisions contain highly variable numbers of spectra, global statistics were computed using weighted averages in order to obtain representative estimates of the effective computational cost over the entire dataset. The reported per-sample costs correspond to the average execution time required to process a single spectrum, computed as

\begin{equation}
\bar{t} =
\frac{\displaystyle\sum_{i=1}^{N} \sum_{j=1}^{8} n_{ij} \, t_{ij}}
     {\displaystyle\sum_{i=1}^{N} \sum_{j=1}^{8} n_{ij}},
\end{equation}

where $i$ indexes the test configurations (combining season, geographical subdivision, and spectral input type), $j$ indexes the soil type ($j = 0, \ldots, 7$), $t_{ij}$ denotes the computational cost per spectrum for configuration $i$ and soil type $j$, and $n_{ij}$ is the corresponding number of processed spectra.

The cumulative total training time is defined as

\begin{equation}
T_{\mathrm{train}} = \sum_{i=1}^{N} \sum_{j=1}^{8} n_{ij} \, t_{ij},
\end{equation}

and provides a direct measure of the overall computational burden associated with the training procedure across the full multi-seasonal dataset.

Statistics were aggregated separately for the CHANNELS and PCA configurations over all seasons, geographical subdivisions, and spectral input types (BT and RAD). The cumulative total training time and the weighted mean and standard deviation of the per-sample prediction cost are reported in Table~7 of the main text.

The moderate overhead observed in the total training time indicates that the dimensionality reduction procedure does not critically affect the overall training efficiency of the SVM classifier, whereas the prediction phase is considerably more computationally demanding for PCA configurations. This behavior is primarily associated with the large effective dimensionality retained by the PCA representation: as reported in Table~S4, the median number of retained principal components ranges from approximately 120 to 350 depending on the season and configuration, with interquartile ranges exceeding 200 components in several cases. Such high-dimensional feature spaces increase the cost of evaluating the Gaussian kernel during classification, which scales with the number of support vectors and the input dimensionality.

This variability should also be interpreted in light of the kernel parametrization strategy adopted in this work and reported in Subsection~\ref{subsec:S_svm_configuration} of the Supplementary Material. Since the Gaussian kernel parameter depends on the dimensionality and statistical properties of the corresponding training representation, the PCA-based configurations are characterized by a kernel evaluation cost that changes from one experiment to another according to the retained number of principal components. By contrast, in the CHANNELS configuration the input dimensionality remains fixed to 45 selected channels, leading to a more stable kernel parametrization and a lower prediction cost.

The memory footprint of the deployed model was evaluated by isolating the components strictly required for inference on new spectra: the trained SVM object, the truncated PCA transformation matrix, and the standardization parameters (mean and standard deviation from the training set). Since the CISVM architecture loads a single soil-type-specific sub-model at a time, according to the surface classification of the processed spectrum, the relevant memory footprint for operational deployment corresponds to the largest individual sub-model rather than the cumulative size across all soil types. This worst-case single-model footprint, consistently associated with the Non-land sub-model, ranges from approximately 170~MB in Summer to approximately 477~MB in Spring, following the same seasonal variation observed for the retained PCA dimensionality (Table~S4). For completeness, we also report the cumulative memory footprint required if all eight soil type sub-models were simultaneously held in memory, e.g., to avoid repeated model loading during processing: even under this more conservative scenario, the total footprint remains modest, ranging from approximately 305~MB in Summer to approximately 747~MB in Spring.

\section{Supplementary Material for Section 5: Comparison with MODIS cloud mask}\label{sec:MODIS_cloud_mask}
Table~\ref{tab:Orbits_test_cloudmask} reports the complete list of the IASI orbits used to construct the collocated datasets for the comparison with MODIS cloud products, described in Section~5 of the main text. For each selected day, the table provides the acquisition date, starting time, and orbit identifier for all consecutive orbits included in the Winter and Summer subsets.

\begin{table}[H]
    \centering
    \resizebox{\textwidth}{!}{%
    \begin{tabular}{|c|c|c|c|c|c|}
        \hline
        \multicolumn{3}{|c|}{\textbf{Winter}} & \multicolumn{3}{c|}{\textbf{Summer}} \\
        \hline
        \textbf{Start Date} & \textbf{Start Time} & \textbf{Orbit ID} 
        & \textbf{Start Date} & \textbf{Start Time} & \textbf{Orbit ID} \\
        \hline
        2022-01-21 & 23:32:55 & 16650 & 2022-07-18 & 23:50:57 & 19179 \\
        \hline
        \textbf{End Date} & \textbf{End Time} & \textbf{Orbit ID} 
        & \textbf{End Date} & \textbf{End Time} & \textbf{Orbit ID} \\
        \hline 
        2022-01-22 & 23:11:59 & 16664 & 2022-07-19 & 23:29:53 & 19193 \\
        \hline
        \multicolumn{3}{|c|}{Total Number of Orbits: 15} & \multicolumn{3}{c|}{Total Number of Orbits: 15} \\
        \hline
    \end{tabular}%
    }
    \caption{Summary of starting and ending IASI orbits used for the test set in Winter and Summer. The orbits are daily and consecutive, spanning an acquisition period of approximately 24 hours for each season.}
    \label{tab:Orbits_test_cloudmask}
\end{table}

\bibliographystyle{elsarticle-num}